\documentclass[onecolumn,showpacs,11pt]{revtex4}
\usepackage{graphicx}
\usepackage{dcolumn}
\usepackage{bm}
\begin{document}
%\preprint{APS/123-QED}
%%%%%%%%%%%%%%%%%%%%%%%%
\newcommand{\hs}{\hspace*{0.5cm}}
\newcommand{\vs}{\vspace*{0.5cm}}
\newcommand{\be}{\begin{equation}}
\newcommand{\ee}{\end{equation}}
\newcommand{\bea}{\begin{eqnarray}}
\newcommand{\eea}{\end{eqnarray}}
\newcommand{\ben}{\begin{enumerate}}
\newcommand{\een}{\end{enumerate}}
\newcommand{\bde}{\begin{widetext}}
\newcommand{\ede}{\end{widetext}}
\newcommand{\nn}{\nonumber}
\newcommand{\crn}{\nonumber \\}
\newcommand{\Tr}{\mathrm{Tr}}
\newcommand{\non}{\nonumber}
\newcommand{\noi}{\noindent}
\newcommand{\al}{\alpha}
\newcommand{\la}{\lambda}
\newcommand{\bet}{\beta}
\newcommand{\ga}{\gamma}
\newcommand{\va}{\varphi}
\newcommand{\om}{\omega}
\newcommand{\pa}{\partial}
\newcommand{\+}{\dagger}
\newcommand{\fr}{\frac}
\newcommand{\bc}{\begin{center}}
\newcommand{\ec}{\end{center}}
\newcommand{\Ga}{\Gamma}
\newcommand{\de}{\delta}
\newcommand{\De}{\Delta}
\newcommand{\ep}{\epsilon}
\newcommand{\varep}{\varepsilon}
\newcommand{\ka}{\kappa}
\newcommand{\La}{\Lambda}
\newcommand{\si}{\sigma}
\newcommand{\Si}{\Sigma}
\newcommand{\ta}{\tau}
\newcommand{\up}{\upsilon}
\newcommand{\Up}{\Upsilon}
\newcommand{\ze}{\zeta}
\newcommand{\ps}{\psi}
\newcommand{\Ps}{\Psi}
\newcommand{\ph}{\phi}
\newcommand{\vph}{\varphi}
\newcommand{\Ph}{\Phi}
\newcommand{\Om}{\Omega}
%%%%%%%%%%%%%%%%%%%%%%%%

\title{3-3-1-1 model for dark matter}

\author{P. V. Dong}
\email {pvdong@iop.vast.ac.vn} \affiliation{Institute of Physics,
Vietnam Academy of Science and Technology, 10 Dao Tan, Ba Dinh, Hanoi, Vietnam}
\author{H. T. Hung}
\email{hthung@grad.iop.vast.ac.vn} \affiliation{Department  of
Physics, Hanoi University of Education 2, Phuc Yen, Vinh Phuc, Vietnam}
\author{T. D. Tham}
\email{tdtham@pdu.edu.vn}
\affiliation{Institute of Physics,
Vietnam Academy of Science and Technology, 10 Dao Tan, Ba Dinh, Hanoi, Vietnam}

\date{\today}

\begin{abstract}
We show that the $SU(3)_C\otimes SU(3)_L\otimes U(1)_X$ (3-3-1) model of strong and electroweak interactions can naturally accommodate an extra $U(1)_N$ symmetry behaving as a gauge symmetry. Resulting theory based on $SU(3)_C\otimes SU(3)_L\otimes U(1)_X\otimes U(1)_N$ (3-3-1-1) gauge symmetry realizes $B-L=-(2/\sqrt{3})T_8+N$ as a charge of $SU(3)_L\otimes U(1)_N$. Consequently, a residual symmetry, $W$-parity, resulting from broken $B-L$ in similarity to $R$-parity in supersymmetry is always conserved and may be unbroken. There is a specific fermion content recently studied in which all new particles that have wrong lepton-numbers are odd under $W$-parity, while the standard model particles are even. Therefore, the lightest wrong-lepton particle (LWP) responsible for dark matter is naturally stabilized. We explicitly show that the non-Hermitian neutral gauge boson $(X^0)$ as LWP cannot be a dark matter. However, the LWP as a new neutral fermion $(N_R)$ can be dark matter if its mass is in range $1.9\ \mathrm{TeV}\leq m_{N_R}\leq 2.5\ \mathrm{TeV}$, provided that the new neutral gauge boson ($Z'$) mass satisfies $2.2\ \mathrm{TeV}\leq m_{Z'}\leq 2.5\ \mathrm{TeV}$. Moreover, the scalar dark matter candidate $(H'\simeq \eta_3)$ which has traditionally been studied is only stabilized by $W$-parity. All the unwanted interactions and vacuums as often encountered in the 3-3-1 model are naturally suppressed. And, the standing issues on tree-level flavor changing neutral currents and CPT violation also disappear.                  
\end{abstract}


\maketitle

\section{\label{intro}Introduction}

One of obvious experimental evidences that we must go beyond the standard model of fundamental particles and interactions is neutrino oscillations \cite{pdg}, which mean that the neutrinos have hierarchical, small masses and mixing. Among the extensions known, the seesaw mechanism~\cite{seesaw1} is perhaps the most natural one for explaining the above problem with the introduction of heavy right-handed
neutrinos ($\nu_R$) or some kind of new neutral fermions ($N_R$). However, while these assumed particles have been not observed it is useful to ask what is their natural origin. They may arise as fundamental constituents in left-right models \cite{lrm} or ${SO}(10)$ unification \cite{so10}. The presence of these particles might also lead to interesting
consequences such as the baryon asymmetry via leptogenesis~\cite{leptog}. In this work, we will show that they can also exist in a gauge model implying to a class of new particles, odd under a parity symmetry responsible for dark matter.  

The approach is based on $SU(3)_C\otimes SU(3)_L\otimes U(1)_X$ gauge symmetry (thus named 3-3-1) in which the last two groups are extended from the electroweak symmetry of the standard model, while the QCD symmetry is retained. The right-handed neutrinos or new neutral fermions may constitute in fundamental lepton triplets/antitriplets of  
${SU}(3)_L$ to complete the representations, \bea 
\left(\begin{array}{c}
\nu_L\\
e_L\\
\nu^c_R\end{array}\right)
\hs \mathrm{or}\hs 
\left(\begin{array}{c}
\nu_L\\
e_L\\
N^c_R\end{array}\right),\nn \eea so-called the 3-3-1 model with right-handed neutrinos \cite{331r} or the 3-3-1 model with neutral fermions~\cite{dongfla}, respectively (see also~\cite{331m} for a variant). In addition, this approach has intriguing features.  The number of fermion families must be an integral multiple of fundamental color number, which is three, in order to cancel ${SU}(3)_L$ anomaly~\cite{anoma}. There are nine flavors of quarks due to the enlarged electroweak gauge symmetry, so the family number must also be smaller than or equal to five to ensure QCD asymptotic freedom. All these result in an exact family number of three, coinciding with the observation~\cite{pdg}. Since the third family of quarks transforms under ${SU}(3)_L$ differently from the first two, this can explain why top quark is so heavy~\cite{longvan}. The extension can also provide some insights of electric charge quantization observed in the nature~\cite{ecq}.

The extended sectors from the standard model in the 3-3-1 models such as scalar, fermion and gauge might by themselves provide dark matter candidates. It is strongly approved by the gauge interactions, minimal Yukawa Lagrangian and scalar potential that normally couple the new concerned particles (similar to the so-called wrong-lepton particles as defined in the following section) in pairs, in similarity to the case of superparticles in supersymmetry. This is automatically resulted from the specific structure of 3-3-1 gauge symmetry by itself \cite{331m,331r}, which is unlike the conclusion in \cite{pires0,pires}. The first attempts in identifying the dark matter candidates of 3-3-1 models have been previously expressed in \cite{tonasse,longlan}, however a strict treatment on their stability issue and relic abundance has been not given. The stabilization of dark matter in the 3-3-1 models due to extra symmetries has been firstly discussed in \cite{pires0,pires}. To this aim, the lepton number symmetry has been imposed in \cite{pires0} so that the lightest bilepton particle could be stabilized. It is noteworthy that all the unwanted interactions of Yukawa Lagrangian and scalar potential (other than the minimal interactions mentioned) explicitly violate the lepton number \cite{lepto331} which are naturally suppressed due to this symmetry (except the coupling of two lepton and one scalar triplets that violates only flavor lepton numbers, but leading to an unrealistic neutrino mass spectrum; however, in our model discussed below this coupling explicitly disappears due to the total lepton number violation). The $Z_2$ as introduced therein can be in fact not needed. An alternative problem encountered is that the lepton number should be also violated due to five-dimensional effective interactions responsible for the neutrino masses as cited therein.   

In \cite{pires}, the bilepton character of new particles has been arranged to be lost, and the lepton number symmetry was no longer to prevent those unwanted interactions from turning on. So, a $Z_2$ symmetry has been included by hand with appropriate $Z_2$ representation assignments in order to eliminate those unwanted interactions, and this symmetry has been regarded as the one for stabilizing the dark matter \cite{pires}. However, since the $Z_2$ that acts on the model multiplets must be spontaneously broken by the Higgs vacuum, there is no reason why the scalar dark matter which carries no lepton number cannot develop a VEV and decay then. Also in \cite{pires}, a continue symmetry called $U(1)_G$ that acts on component particles, not commuting with the gauge symmetry like the lepton charge before, has been introduced to be equivalently used instead of the $Z_2$ in interpreting the dark matter, getting some reason from the gauge interactions. Let us remind the reader that these interactions of gauge bosons with fermions, scalars or gauge self-interactions are automatic consequences of and already restricted by the gauge symmetry by itself as mentioned. They always present, not excluded or added by other interactions in any cases that the $U(1)_G$ or even lepton charge is imposed or not. For the purpose in suppressing those unwanted interactions and vacuums, obviously there are many other symmetries behaving like $U(1)_G$ or lepton charge found out as respective solutions of the gauge interactions' conservation. However, all these continue symmetries can face problems below apart from their nature of presence.  

The continue symmetries above must be supposed as exact symmetries responsible for the dark matter stability. Therefore, they can be naturally regarded as respective residual symmetries of higher symmetries that span the 3-3-1 group (since they do not commute with the gauge symmetry as mentioned) acting at the Lagrangian level, under which the unwanted interactions are explicitly suppressed. In other words, the minimal Lagrangian of the theory actually contain larger symmetries spanning the gauge symmetry that shall be spontaneously broken down to those residual symmetries, respectively. As a specific property of the 3-3-1 models, the lepton charge (or even any kind of $U(1)_G$ if one independently includes, neglected of the lepton charge symmetry) should work as the residual gauge symmetry of some higher symmetry mentioned (as shown in the next section) and must be spontaneously broken so that the resulting gauge boson gets a large enough mass to make a consistency of the theory. On the other hand, this lepton charge symmetry or even a general continue symmetry mentioned is also known to be actually violated due to its anomalies. Therefore, such symmetries (lepton number or $U(1)_G$) would be no longer to protect the dark matter stability from decays. For the stability issue of dark matter, similarly to the $R$-parity in supersymmetry it is more natural to search for an exact and unbroken residual discrete symmetry of some anomaly-free continue symmetry spanning either the lepton number and other necessary symmetries such as baryon charge or some kind like $U(1)_G$. Let us remark that among the continue symmetries analyzed, the one concerning lepton charge is perhaps the most motivative and natural because of the followings: (i) all the unwanted interactions in ordinary 3-3-1 models which should be prevented in fact violate the lepton numbers \cite{lepto331}; (ii) while the discrete symmetry is conserved responsible for dark matter stability, the lepton or baryon numbers could be broken in several ways necessarily to account for the observed neutrino masses and baryon-number asymmetry. By the way, we will see that this is similar to enlarge the $SU(5)$ theory to $SO(10)$ in which the $B-L$ charge is naturally gauged. In this work the lepton charge will be taken into account which is different from the $U(1)_G$ symmetry ad hoc input.                                      

By investigating of nontrivial lepton number behavior and a resulting $W$-parity (similar to the $R$-parity in supersymmetry) in a specific 3-3-1 model \cite{dongfla}, we show that the theory can contain natural dark matter candidates. For details, we consider the 3-3-1 model with neutral fermions~($N_R$) which is different from the model of \cite{pires}. These neutral fermions possess no lepton number as already studied before in a TeV seesaw extension of the standard model \cite{ma} and in the 3-3-1 model with flavor symmetries \cite{dongfla}. We investigate lepton number symmetry, its dynamics and other symmetries which result in a new 3-3-1-1 gauge model. We show that there is an unbroken residual symmetry of such (anomaly-free) 3-3-1-1 theory behaving like the $R$-parity in supersymmetry under which almost the new particles as given are odd. It is interesting that the model can contain several kinds of dark matter such as singlet scalar, fermion and gauge boson as often presented in other extensions of the standard model and similarity to the conclusion in \cite{pires}. However, these candidates may be heavy as some TeV which is unlike light candidates in the standard model familiar extensions. Before \cite{pires} and our work, the previous considerations of the 3-3-1 model dark matters recognized only scalar singlet \cite{longlan} and lightest supersymmetric particles of respective supersymmetric 3-3-1 versions. The reason behind this may be that the mentioned symmetries such as 3-3-1-1 and $W$-parity under which the dark matters are dynamically stabilized have been not explored yet. The dark matter phenomenologies in our models will be different from the other extensions. The model can work better under the experimental constraints than the ordinary 3-3-1 model with right-handed neutrinos due to the $W$-parity mentioned. The above procedure will fail when applying for the other 3-3-1 models such as the model of \cite{pires}, the 3-3-1 model with right-handed neutrinos \cite{331r} and the minimal 3-3-1 model \cite{331m}. In fact, all the particles including the new ones in those models would transform trivially under the $W$-parity. Therefore, this parity may be only realized in the class of 3-3-1 models with flavor symmetries \cite{dongfla}.  

The rest of this article is organized as follows. In Sec. \ref{dmodel} we give a review of the 3-3-1 model with neutral fermions by stressing on baryon and lepton numbers as well as proposing of wrong-lepton particles. We next construct the 3-3-1-1 gauge symmetry, $W$-parity, and showing possible dark matter candidates and its direct consequences. We also identify physical scalar particles and giving a discussion of all the masses. In Sec. \ref{modelrelic} we provide detailed calculations of relic densities of possible dark matters and showing constraints. Finally, we summarize our results and making outlooks in the last section -- Sec. \ref{con}.         

\section{\label{dmodel}Proposal of 3-3-1-1 model}

\subsection{The 3-3-1 model with neutral fermions and wrong-lepton particles}

The gauge symmetry of the 3-3-1 model under consideration is given by 
${SU}(3)_C\otimes {SU}(3)_L \otimes {U}(1)_X$, where the first factor is the usual QCD symmetry while the last two ${SU}(3)_L \otimes {U}(1)_X$ are extended from the electroweak symmetry of the standard model. The electric charge operator is the only unbroken residual charge of the gauge symmetry and being defined by
$Q=T_3-({1}/{\sqrt{3}})T_8+X$, where $T_i$ $(i=1,2,3,...,8)$ are the ${SU}(3)_L$ charges while $X$ is that of $U(1)_X$. The weak hypercharge of the standard model is therefore identified as $Y=-({1}/{\sqrt{3}})T_8+X$.  

The fermion content which is anomaly free is assigned by 
\bea \psi_{aL} &=& \left(\begin{array}{c}
               \nu_{aL}\\ e_{aL}\\ (N_{aR})^c
\end{array}\right) \sim (1,3, -1/3),\hs e_{aR}\sim (1,1, -1),
\\
Q_{\al L}&=&\left(\begin{array}{c}
  d_{\al L}\\  -u_{\al L}\\  D_{\al L}
\end{array}\right)\sim (3,3^*,0),\hs Q_{3L}=\left(\begin{array}{c} u_{3L}\\  d_{3L}\\ U_L \end{array}\right)\sim
 \left(3,3,1/3\right),\\ u_{a
R}&\sim&\left(3,1,2/3\right),\hs d_{a R} \sim
\left(3,1,-1/3\right),\\ U_{R}&\sim& \left(3,1,2/3\right),\hs D_{\al R}
\sim \left(3,1,-1/3\right),\eea  where $a = 1, 2, 3$ and $\al=1,2$ are family indices. The values defined in parentheses are quantum numbers based on $({SU}(3)_C, {SU}(3)_L, {U}(1)_X)$ gauge symmetries, respectively. The $N_{aR}$ and $U,\ D_\al$ are the new neutral fermions (which are singlets under the standard model symmetry as the right-handed neutrinos often considered) and exotic quarks, respectively. The electric charges of exotic quarks $Q(U)=2/3$ and $Q(D_\al)=-1/3$ are the same ordinary quarks. As mentioned, the lepton number of $N_{aR}$ will be taken to be zero: $L(N_{aR})=0$. This is due to the fact that the conventional seesaw mechanism with right-handed neutrinos including that of the 3-3-1 model can require a very high seesaw scale of $M_R\sim 10^{10}-10^{14}$ GeV \cite{seesaw1,331seesaw}. As already shown in \cite{ma,dongfla}, if such $N_R$ supposed one can have a natural TeV seesaw scale matching the 3-3-1 breaking scale. And, the lepton mixing matrix under flavor symmetries can be naturally explained in those models \cite{ma1,dongfla}. The presence of $N_R$ also implies to a kind of new particles that is odd under a parity symmetry, well-motivated responsible for dark matter candidates as shown below.       

To break the gauge symmetry and generating the masses, this kind of the 3-3-1 model actually requires three scalar triplets \cite{331r}
\bea
\rho &=& \left(\begin{array}{c}
\rho^+_1\\
\rho^0_2\\
\rho^+_3\end{array}\right)\sim (1,3,2/3),\label{vev2}\\ 
\eta &=&  \left(\begin{array}{c}
\eta^0_1\\
\eta^-_2\\
\eta^0_3\end{array}\right)\sim (1,3,-1/3),\label{vev1}\\
 \chi &=& \left(\begin{array}{c}
\chi^0_1\\
\chi^-_2\\
\chi^0_3\end{array}\right)\sim (1,3,-1/3).\label{vev3}\eea Here, the gauge symmetry is broken via two stages: the first stage ${SU}(3)_L \otimes {U}(1)_X$ is broken down to that of the standard model generating the masses of new particles such as exotic quarks $U$, $D_\al$ as well as the new gauge bosons: one neutral $Z'$ coupled to the generator that is orthogonal to the weak hypercharge and two charged $X^{0/0*}$, $Y^\mp$ corresponding to $T_4\pm i T_5$ and $T_6\pm i T_7$ raising/lowering operators. In the second stage the standard model electroweak symmetry is broken down to ${U}(1)_Q$ responsible for the masses of ordinary particles such as $W^\pm$, $Z$, $e_a$, $u_a$, and $d_a$.
 
The lepton number $(L)$ of lepton triplet components is given by $(+1,+1,0)$ which does not commute with the $SU(3)_L$ gauge symmetry unlike the standard model case:
\be [L,T_4\pm i T_5]=\pm(T_4\pm i T_5)\neq 0,\hs [L,T_6\pm i T_7]=\pm(T_6\pm i T_7)\neq 0,\ee which mean that $X$ and $Y$ bosons carry lepton numbers with one unit. This also happens for the 3-3-1 model with right-handed neutrinos and the minimal 3-3-1 model. It is a characteristic property of this kind of the model. Hence, if the lepton number is a symmetry of the theory, it can be regarded as a residual charge of conserved symmetries, \be \mathcal{G}\equiv SU(3)_{L}\otimes {U}(1)_{\mathcal{L}},\ee where the second factor is a new symmetry supposed because the lepton number and the gauge symmetry generators do not form a closed algebra. [This is because in order for $L$ to be some generator of $SU(3)_L$, the $L$ charges of a complete multiplet must add up to zero which is not correct. Also, the $\mathcal{L}$ and $X$ charges must be distinct because for the $SU(3)_L$ fermion singlets, the lepton number and electrical charge generally do not coincide]. The lepton number that is a combination of $SU(3)_{L}\otimes {U}(1)_{\mathcal{L}}$ diagonal generators (due to the conservation of lepton number as strictly required by the experiments, in similarity to the case of electric charge operator) can be easily obtained by acting it on a lepton triplet to be given by \be L=\fr{2}{\sqrt{3}}{T}_8+\mathcal{L},\label{dong111}\ee where the ${T}_8$ is the charge of $SU(3)_{L}$ while the $\mathcal{L}$ is that of $U(1)_{\mathcal{L}}$~\cite{lepto331}. On the grounds of known lepton numbers, the $\mathcal{L}$ charges of all the model multiplets can be easily obtained as given in Table~\ref{tb1} (notice that the lepton numbers of $\chi^0_3$, $\rho^0_2$ and $\eta^0_1$ must be zero since these fields are normally required to develop VEVs responsible for the gauge symmetry breaking and mass generation).
\begin{table}[htdp]
\begin{center}
\begin{tabular}{|c|ccccccccccc|}
\hline
Multiplet & $\psi_{aL}$ & $e_{aR}$ & $Q_{3L}$ & $Q_{\al L}$ & $u_{a R}$ & $d_{a R}$ & $U_R$ & $D_{\al R}$ & $\rho$ & $\eta$ & $\chi$ \\ \hline
$\mathcal{L}$ & $\fr 2 3$ & $1$ & $-\fr 1 3$& $\fr 1 3$ & 0& 0 & $-1$ & 1& $-\fr 1 3$ & $-\fr 1 3$ & $\fr 2 3 $ \\ \hline
\end{tabular}
\end{center}
\caption{\label{tb1} The $\mathcal{L}$-charge of model multiplets.}
\end{table}%
Moreover, it is able to point out that all the ordinary interactions of the theory (i.e. the minimal interactions as mentioned) conserve $\mathcal{L}$ \cite{lepto331}. For a convenience in reading, we give also the lepton numbers of model particles $L$ in Table~\ref{tb2}.
\begin{table}[htdp]
\begin{center}
\begin{tabular}{|c|cccccccccccccccccccccc|}
\hline
Particle & $\nu_{aL}$ &  $e_a$ & $N_{aR}$ & $u_a$ & $d_a$  & $U$ & $D_\al$ & $\rho^+_1$ & $\rho^0_2$ & $\rho^+_3$ & $\eta^0_1$ & $\eta^-_2$ & $\eta^0_3$ & $\chi^0_{1}$ & $\chi^-_2$ & $\chi^0_3$ & $\gamma$ & $W$ & $Z$ & $Z'$ & $X^0$ & $Y^-$ \\ \hline
$L$ & 1 & 1 & 0 & 0 & 0 & $-1$ & 1 & 0 & 0 & $-1$ & 0 & 0 & $-1$ & 1 & $1$ & 0 & 0 & 0 & 0 & 0 & 1 & 1 \\
\hline
\end{tabular}
\end{center}
\caption{\label{tb2} The lepton number of model particles.}
\end{table}
We see that the standard model particles have lepton numbers as usual. However, almost the new particles such as $N_R$, $U$, $D$, $X$, $Y$, $\rho_3$, $\eta_3$, and $\chi_{1,2}$ possess unnormal lepton numbers in comparison to those of the standard model nature. For example, should $L(U,D)=0$ like ordinary quarks instead of $\pm 1$. This kind of particles is going to be named as ``wrong-lepton particles'' or sometimes ``$W$-particles'' for short.  
 
In this work, we suppose that the $\mathcal{G}$ symmetry and thus $U(1)_{\mathcal{L}}$ responsible for the lepton number is an exact symmetry. However, since the scalar triplets as given are all charged under $\mathcal{G}$ and will get VEVs, the $\mathcal{G}$ symmetry must be broken spontaneously (in accompany with the gauge symmetry breaking). It is also easily shown that the scalar potential can be stabilized by the following solution of the potential minimization conditions:  
\bea \langle \chi^0_1\rangle &=& \langle \eta^0_3\rangle =0,\label{vev01}\\ 
\langle \rho^0_2\rangle &\neq& 0,\hs \langle \eta^0_1\rangle\neq 0,\hs \langle \chi^0_3\rangle \neq 0.\label{vev02}\eea In fact, this solution of the potential minimization has been formally used in the literature as a standard vacuum structure \cite{331r}. Also, with the VEVs as given in (\ref{vev01}) and (\ref{vev02}), i.e.
\bea \langle \rho \rangle = \fr{1}{\sqrt{2}}(0,v,0)^T,\hs \langle \eta \rangle =\fr{1}{\sqrt{2}}(u,0,0)^T,\hs \langle \chi \rangle =\fr{1}{\sqrt{2}}(0,0,\om)^T,\eea
all the particles in this model (except for $\nu_L$ and $N_R$) will get consistent masses at the tree level in similarity to those of the ordinary 3-3-1 model with right-handed neutrinos \cite{331r}. Let us note that $\om$ is responsible for the first stage of electroweak symmetry breaking $SU(3)_L\otimes U(1)_X\longrightarrow SU(2)_L\otimes U(1)_Y$ providing the masses for new particles, whereas $u,v$ act on the second stage $SU(2)_L\otimes U(1)_Y\longrightarrow U(1)_Q$ giving the masses for ordinary particles. To keep a consistency with the standard model, we should suppose \be u^2,\ v^2\ll \om^2.\ee For the $\mathcal{G}$ symmetry, although both the $\mathcal{L}$ and $T_8$ (and all other generators) are broken, the combination of lepton number $L$ in this case is conserved by the VEVs which can be verified directly from Table~\ref{tb2}. The brokendown of the $\mathcal{G}$ symmetry into the lepton number \be \mathcal{G}=SU(3)_{L}\otimes U(1)_{\mathcal{L}}\longrightarrow U(1)_L\ee implies the existence of eight Goldstone bosons contained in the scalar sector $\rho,\ \eta,\ \chi$. However, these Goldstone bosons are just those associated with the gauge symmetry breaking $SU(3)_L\otimes U(1)_X\longrightarrow U(1)_Q$ that will be subsequently gauged away (they are unphysical because they are already the Goldstone bosons of the gauge symmetries as stated).     

Moreover, by a similar ingredient the baryon number ($B$) may be found to be not commuted with the gauge symmetry as well as being resulted from some broken exact symmetries, \be\mathcal{G}'\equiv SU(3)_L\otimes U(1)_{\mathcal{B}}\longrightarrow U(1)_B,\ee because the baryon numbers of $U,\ D_\al$ are unknown and in principle could be arbitrary (in this case the unwanted interactions also violate the baryon number and being suppressed due to the $\mathcal{B}$ charge conservation). The $\chi^0_1$ and $\eta^0_3$ will carry the baryon number in this case which can only be conserved by the above potential minimization condition. For a simplicity, in this work let us take $B(U)=B(D_\al)=1/3=B(u_a)=B(d_a)$ and vanishing for other fields as actually used in the literature for this kind of the model \cite{lepto331} so that \be B=\mathcal{B}\label{sobari}\ee of the theory commuting with the gauge symmetry and being always conserved at the renormalizable level as in the standard model (i.e., the baryon number of our general theory is always an exact and unbroken symmetry since the unwanted interactions conserve $B$ while the $\chi^0_1$ and $\eta^0_3$ are neutral under this charge). The $\mathcal{B}$ charges of model multiplets are given by Table~\ref{tb4}.
\begin{table}[htdp]
\begin{center}
\begin{tabular}{|c|c|c|c|c|c|c|c|c|c|c|c|}
\hline
Multiplet & $\psi_{aL}$ & $e_{aR}$ & $Q_{3L}$ & $Q_{\al L}$ & $u_{a R}$ & $d_{aR}$ & $U_R$ & $D_{\al R}$ & $\rho$ & $\eta$ & $\chi$ \\ 
\hline
$\mathcal{B}$ & 0 & 0 & $\fr 1 3$ & $\fr 1 3$ & $\fr 1 3$ & $\fr 1 3$ & $\fr 1 3$ & $\fr 1 3$ & 0 &0 &0 \\
\hline  
\end{tabular}
\end{center}
\caption{\label{tb4} The $\mathcal{B}$-charge of model multiplets.}
\end{table}
[Let us remark on alternative cases: (i) If there was $\langle \chi^0_1\rangle \neq 0$ or $\langle \eta^0_3\rangle \neq 0$, the $L$ would be broken too, along with $T_8$ and $\mathcal{L}$. (ii) The conservation of $L$ in this model is not an automatic consequence of the theory like the standard model. This is because if the $U(1)_{\mathcal{L}}$ symmetry was not imposed there would be the unwanted interactions explicitly violating $L$ as actually seen in the Yukawa sector and/or scalar sector \cite{331seesaw,lepto331}. (iii)  If the baryon numbers of $U,\ D_\al$ were alternatively chosen, the (i) and (ii) would also apply for the baryon number.]

The above ingredients of lepton and baryon numbers have been presented only for the 3-3-1 model with neutral fermions. In general, it can be also applied for the minimal 3-3-1 model \cite{331m} and 3-3-1 model with right-handed neutrinos \cite{331r}. Here the crucial discrimination is that in these models wrong-lepton particles differ from the ordinary ones by two units in lepton charge which have been also called as bilepton particles, whereas in the present model it differs only one unit due to a possible nature of the neutral fermions $N_R$. However, they are completely distinguished when replying to the dark matter problems as shown below.       

\subsection{3-3-1-1 gauge symmetry and $W$-parity}

Let us recall that the lepton numbers $L$ and $\mathcal{L}$ which satisfy (\ref{dong111}) were primarily introduced in other 3-3-1 models \cite{lepto331}, however their dynamical nature has been completely not realized and examined yet. In the second article of \cite{dongfla}, we have given the first notes on this lepton dynamics. And, in the current work it is to be analyzed in more details. It is the fact that since the $T_8$ is a gauged charge of the $SU(3)_L$ symmetry the $L$, thus the $\mathcal{L}$, and vise versa must be subsequently gauged. This is because by a contrast assumption that both the $L$ and $\mathcal{L}$ are global generators, the $T_8\sim L-\mathcal{L}$ is also global which is incorrect. In this case, the anomalies coupled to $L$, thus to $\mathcal{L}$, are obviously unable to suppress, which spoils the model's consistency. To regard the lepton number as such a local symmetry for this kind of the model (or leptonic dynamics is considered) we must deal with the leptonic anomaly cancellation issue. All this also applies for the baryon number if it satisfies (iii); however it does not by our choice. 

A solution to canceling the leptonic anomalies is that we can consider the model with gauged symmetry $N\equiv \mathcal{B}-\mathcal{L}$ (where $\mathcal{B}$ is the baryon number as given above) as well as introducing three new right-handed neutrinos transforming as singlets under the 3-3-1 symmetry,
\be \nu_{aR}\sim (1,1,0).\ee Here, these particles which have lepton and baryon numbers as usual, $\mathcal{L}(\nu_{aR})=L(\nu_{aR})=1$ and $\mathcal{B}(\nu_{aR})=B(\nu_{aR})=0$, are necessarily included in order to cancel the gravitational anomaly $[\mathrm{Gravity}]^2U(1)_N$ (since this anomaly of $\nu_L$ and $N_R$ is not cancelled out). It is explicitly checked that the resulting theory is free from all the anomalies as presented in Appendix \ref{appendixa}. Hence, the new theory as an important investigation of this article is obtained by the following gauge symmetry: 
\be {SU}(3)_C\otimes {SU}(3)_L \otimes {U}(1)_X \otimes U(1)_N,\ee so called the 3-3-1-1 model. And, the multiplets of the 3-3-1-1 model as well as their $N$-charges can be easily counted to be given in Table \ref{bangn}. Here, the complex scalar 3-3-1 singlet, \be \phi\sim(1,1,0),\ee with $\mathcal{B}(\phi)=B(\phi)=0,\ \mathcal{L}(\phi)=L(\phi)=-2$ has been included along with $\eta,\ \rho,\ \chi$ for breaking the 3-3-1-1 symmetry and generating the masses in a correct way.  
\begin{table}[htdp]
\begin{center}
\begin{tabular}{|c|c|c|c|c|c|c|c|c|c|c|c|c|c|}
\hline
Multiplet & $\psi_{aL}$ & $e_{aR}$ & $\nu_{aR}$ & $Q_{3L}$ & $Q_{\al L}$ & $u_{a R}$ & $d_{aR}$ & $U_R$ & $D_{\al R}$ & $\rho$ & $\eta$ & $\chi$ & $\phi$ \\ 
\hline
$N=\mathcal{B-L}$ & $-\fr 2 3$ & $-1$ & $-1$ & $\fr 2 3$ & $0$ & $\fr 1 3$ & $\fr 1 3$ & $\fr 4 3$ & $-\fr 2 3$ & $\fr 1 3$ & $\fr 1 3$ & $-\fr 2 3$ & $2$ \\
\hline  
\end{tabular}
\end{center}
\caption{\label{bangn} The 3-3-1-1 model multiplets and respective $N$-charges.}
\end{table}
Let us remind the reader that the $B-L$ gauge charge, which can be directly derived from (\ref{dong111}) and (\ref{sobari}) as follows \be B-L=-\fr{2}{\sqrt{3}}T_8+N,\ee is a residual symmetry of $SU(3)_L\otimes U(1)_N$ that does not commute with the 3-3-1 gauge symmetry in similarity to the lepton charge $L$. The extension from the 3-3-1 gauge symmetry to the new 3-3-1-1 gauge symmetry which must also apply for the ordinary 3-3-1 models that respect the lepton number symmetry by this view is very intriguing and quite similar as enlarging the $SU(5)$ theory to $SO(10)$ in which the $B-L$ charge is naturally gauged. 

While this possibility of a phenomenological 3-3-1-1 model is interesting and worth exploring to be published elsewhere~\cite{dongbl331}, in this work we will focus on only its consequence of a discrete residual symmetry responsible for the dark matter stabilization as shown below. Therefore, the leptonic and baryonic dynamics as associated with the new gauge charge $N$ will be neglected. The lepton number will be understood as a consequence of the charge conservation associated with $\mathcal{G} = SU(3)_{L}\otimes U(1)_{\mathcal{L}}$ symmetry in which the first factor is a global version of the gauge symmetry. [Namely, in calculating lepton numbers all $SU(3)_L$ global quantum numbers for model multiplets are the same gauged ones. And, both ${T}_8$ and $\mathcal{L}$ in case responsible for the lepton number will be taken as global charges and not gauged, which should be not confused to the similar ones of the $SU(3)_L\otimes U(1)_X$ gauge symmetry.] Similarly, the baryon number $B$ will be regarded as an ordinary global charge. Since the general theory is always conserved by the baryon number, talking about $\mathcal{L}$ or $L$ is equivalent to the $N$ charge which should be understood in the following discussions. In other words, the 3-3-1 model with neutral fermions and $\mathcal{L}$-charge (plus the new right-handed neutrinos and scalar singlet) is also understood as the 3-3-1-1 model in which the gauge interactions or dynamics as associated with $N$-charge (thus $B$ and $L$) is omitted in this work.      
   
Although the $U(1)_{\mathcal{L}}$ and $U(1)_{\mathcal{B}}$ symmetries have been imposed and $L,\ B$ being conserved by the VEVs of $\eta,\ \rho,\ \chi$ as given, it is evident that $B,\ L$ should be broken in some way in order to account for the matter-antimatter asymmetry of the universe and even neutrino masses included later. On the other hand, as stated the nature of $L$ in this model is a gauged charge since it is a result from the~$T_8$. The theory with $L$ gauged simplest takes $N=\mathcal{B-L}$ into account since this new charge is necessarily independent of the anomalies, and the complete brokendown of the $N$ charge must be achieved so that its gauge boson $Z_{N}$ gets a large enough mass to escape from the current detectors. All these can be achieved by the scalar singlet $\phi$ when it develops a VEV,\be \langle \phi\rangle =\fr{1}{\sqrt{2}}\La. \ee Therefore, we will assume that the matter parity (quite similar to the MSSM case), a residual discrete symmetry of broken $B-L=-(2/\sqrt{3})T_8+N$ [or $SU(3)_L\otimes U(1)_N$], thus $R$-parity when included spin is an exact and unbroken symmetry of the 3-3-1-1 theory: 
\be P=(-1)^{3(B-L)+2s}=(-1)^{-2\sqrt{3}T_8+3N+2s},\ee which still conserves all the vacuum structures above. (For a detailed proof, see Appendix \ref{appendixb}.) The $R$-parity of model particles is given in Table~\ref{tb3}.
\begin{table}[htdp]
\begin{center}
\begin{tabular}{|l|l|}
\hline
$+1$ (ordinary or bilepton particle) & $\nu_L$\hs $e$\hs  $u$\hs  $d$ \hs $\gamma$ \hs  $W$\hs  $Z$\hs $\rho_{1,2}$\hs  $\eta_{1,2}$\hs  $\chi_3$\hs $\phi$\hs $\nu_{R}$\hs $Z'$\hs $Z_N$ \\
\hline
$-1$ (wrong-lepton particle) & $N_R$\hs $U$\hs $D$\hs $\rho_3$\hs $\eta_3$\hs $\chi_{1,2}$\hs $X$\hs $Y$ \\ \hline
\end{tabular}
\end{center}
\caption{\label{tb3} The $R$-parity of 3-3-1-1 model particles that separates wrong-lepton particles from ordinary particles. Here the family indices for fermions have been suppressed and should be understood.}
\end{table} We see that all the particles having unusual (unnormal) characteristic lepton-number differing from the ordinary one by one unit, e.g. $L(N_R)=0$, $L(U)=-1$, $L(X)=1$, $L(\rho_3)=-1$, as already named wrong-lepton particles, are odd. Otherwise the ordinary particles such as the standard model particles or new particles that remain their would-be-ordinary properties of the lepton number (or differing from the ordinary ones by an even lepton number as $\phi$ due to the parity symmetry) are even. It is remarkable that the $R$-parity $(P)$ which originates in the 3-3-1-1 gauge symmetry is a natural symmetry of wrong-lepton particles in this model. The lightest wrong-lepton particle (LWP) within the odd ones is stable and able to contribute to dark matter since this parity is exact, not broken by the VEVs. Simultaneously, as mentioned we can have several violations of $L$ or $B$ (one example is that $L=\pm 2$ is broken by $\La$) in order to make the model all phenomenologically viable while still protect the parity. Under this view, it is noteworthy that there can be $R$-parity odd particles, wrong-lepton particles, even in nonsupersymmetric theories like ours. This is due to a possible property of neutral fermions $L(N_R)=0$ as also implemented by a class of 3-3-1 models with flavor symmetries \cite{dongfla}. By contrast, every particle in the 3-3-1 model with right-handed neutrinos $L(\nu_R)=1$ and the minimal 3-3-1 model is even under the parity.
   
Also, it can be explicitly pointed out that in the interactions of theory all the odd particles are only coupled in pairs, so linked to ordinary particles of the standard model due to the 3-3-1 gauge symmetry, the $U(1)_{\mathcal{L}}$ symmetry and the vacuum respecting $R$-parity (i.e. the 3-3-1-1 symmetry with conserved $R$-parity). Let us show now for examples and consequences (the scalar potential also possessing these properties to be expressed latter): 
\ben \item Yukawa sector
\bea \mathcal{L}_{Y}&=&h^e_{ab}\bar{\psi}_{aL}\rho e_{bR} +h^\nu_{ab}\bar{\psi}_{aL}\eta\nu_{bR}+h'^\nu_{ab}\bar{\nu}^c_{aR}\nu_{bR}\phi + h^U\bar{Q}_{3L}\chi U_R + h^D_{\al \beta}\bar{Q}_{\al L} \chi^* D_{\beta R}\crn 
&&+ h^u_a \bar{Q}_{3L}\eta u_{aR}+h^d_a\bar{Q}_{3L}\rho d_{aR} + h^d_{\al a} \bar{Q}_{\al L}\eta^* d_{aR} +h^u_{\al a } \bar{Q}_{\al L}\rho^* u_{aR}\crn
&&+H.c.\eea 
We see the odd scalars $\rho_3$, $\eta_3$ and $\chi_{1,2}$ do not interact with ordinary fermions which only couple to $eN$, $\nu N$, $uU$, $dU$, $dD$, $uD$ with the type of an even-odd particle pair due to the 3-3-1 symmetry. There are not $\mathcal{L}$-charge violating similar interactions (which lead to violations of $R$-parity) such as $\bar{\psi}_{aL}\chi \nu_{bR}$, $\bar{\psi}^c_{aL}\psi_{bL}\rho$, $\bar{Q}_{3L}\chi u_{aR}$, $\bar{Q}_{\al L} \chi^* d_{aR}$, $\bar{Q}_{3L}\eta U_R$ and so forth. In addition, since $R$-parity is conserved the VEVs of $\eta^0_3$ and $\chi^0_1$ vanish. Due to these conditions, the ordinary quarks and exotic quarks do not mix which means that the flavor changing neutral currents at the tree level disappear. The 3-3-1 model with right-handed neutrinos as often considered~\cite{331r} is strictly improved by this parity. It is also noted that $N_R$ do not mix with $\nu_{L}$ and $\nu_R$ due to the parity symmetry.      
\item Gauge boson sector: The odd gauge bosons $X,\ Y$ do not couple to the standard model gauge boson pairs also, except for a similar type as mentioned such as $WX$, $WY$ or other types as $W$-$W$-odd-odd, etc due to the 3-3-1 symmetry. This can be verified directly from \cite{longsoa}. Due to the $R$-parity symmetry, the neutral gauge boson $X$ cannot mix with $Z$ and $Z'$ bosons. The CPT violation at the tree level as stated in \cite{huyendl} is suppressed. Again, constraints on the 3-3-1 model with right-handed neutrinos \cite{331r} are improved by the parity.        
\een        

We notice that in the MSSM, the spins or angular-momenta of component particles within a supermultiplet do not commute with supersymmetry (comparing to our case in an alternative scenario, the lepton number is not commuted with the gauge symmetry). However, the residual $R$-parity of spin, lepton and baryon numbers (which must also be not commuted with the supersymmetry) is conserved and unbroken, even though the conservation of its global symmetry (that spans such spin, lepton and baryon numbers known as $R$-symmetry) is actually broken along with the supersymmetry breaking. It is also emphasized that the lepton and baryon number conservation of the MSSM superpotential is not an automatic consequence of the theory at the renormalizable level unlike the standard model, which must be an assumption in similarity to our case with the $U(1)_{\mathcal{L}}$ symmetry associated with the lepton number. Our $R$-parity obviously has a different origin of the 3-3-1-1 gauge symmetry as mentioned. This is due to the nature of the lepton charge nontrivially resided in $SU(3)_{L}\otimes U(1)_{\mathcal{L}}$, baryon charge $U(1)_{\mathcal{B}}$ and spin uniform for both kinds of respective particles (ordinary and $W$-particles), instead of those in the MSSM. Particularly, the $\mathcal{G}$ symmetry or $U(1)_{\mathcal{L}}$ of lepton number is broken due to the gauge symmetry breaking. The $L$-symmetry conservation for this model can be also protected by the $R$-parity instead. The breaking of $N=\mathcal{B-L}$ gauge symmetry can happen just above TeV scale or at a very high scale. Therefore, to mark such a different origin we are going to call the $R$-parity in this model to be $W$-parity where $W$ means the item ``wrong-lepton'' as already pointed out. 

Depending on the parameter space of the model, the LWP may belong to the nature of a vector particle ($X^0$), scalar ($\chi^0_1$ or $\eta^0_3$), or fermion ($N^0_R$), which must be considered here to be electrically-neutral if they are expected to contribute the dark matter. Before considering those cases detailed in the next section, let us take some comments on the scalar particle identifications and the masses of particles with stressing on those of the wrong-lepton particles.  

\subsection{Scalar potential and masses}

If the scalar singlet $\phi$ which is responsible for completely breaking $U(1)_N$ as mentioned lives in the same scale of 3-3-1 breaking ($\La\sim \om$), it will couple to ordinary scalars $\eta,\ \rho,\ \chi$ via the scalar potential. And, the phenomenologies associated with broken $B-L$ symmetry via $Z_N$ happen simultaneously with the new physics of the 3-3-1 model just in TeV scale. This possibility is very interesting to be studied \cite{dongbl331}. Otherwise, the $\phi$ should be very massive that can be integrated out from the low-energy effective potential of $\eta,\ \rho,\ \chi$. Also, the $Z_N$ will be decoupled from the gauge boson spectrum. This is the case considered in this work. It is to be noted that the behavior of $W$-parity by both cases is unchanged. For the $\phi$, we can expand \be \phi =\fr{1}{\sqrt{2}}(\La + R+i I).\ee The imaginary part ($I$) of $\phi$ is just the Goldstone boson of $Z_N$, while the real part ($R$) is a new physical Higgs boson carrying a lepton number of two units and being $W$-parity even. The mass of $Z_N$ is proportional to the VEV $\La$ of $\phi$ scalar.       

At the low-energy, the scalar potential of $\eta,\ \rho,\ \chi$ after integrating out the $\phi$ that must conserve the 3-3-1 symmetry and $W$-parity is given by 
\bea V &=& \mu^2_1\rho^\dagger \rho + \mu^2_2 \chi^\dagger \chi + \mu^2_3 \eta^\dagger \eta + \la_1 (\rho^\dagger \rho)^2 + \la_2 (\chi^\dagger \chi)^2 + \la_3 (\eta^\dagger \eta)^2\crn
&&+ \la_4 (\rho^\dagger \rho)(\chi^\dagger \chi) +\la_5 (\rho^\dagger \rho)(\eta^\dagger \eta)+\la_6 (\chi^\dagger \chi)(\eta^\dagger \eta)\crn
&& +\la_7 (\rho^\dagger \chi)(\chi^\dagger \rho) +\la_8 (\rho^\dagger \eta)(\eta^\dagger \rho)+\la_9 (\chi^\dagger \eta)(\eta^\dagger \chi)\crn
&& + (f\epsilon^{mnp}\eta_m\rho_n\chi_p+H.c.), \eea where $\mu_{1,2,3}$ and $f$ have mass dimension whereas $\la_{1,2,3,...,9}$ are dimensionless. The unwanted terms such as $\eta^\dagger \chi$, $(\eta^\dagger \chi)^2$, $(\rho^\dagger\rho) (\eta^\dagger \chi$) and so on which violate $L$ (or $\mathcal{L}$) are prevented due to the parity symmetry. Let us note that the $f$ coupling conserves all the natural symmetries of the theory as imposed and there is no reason why it is not presented. In the literature, it was ordinary excluded \cite{longsca} (see also the first one of \cite{longlan}). Therefore, it is needed to clarify that its presence makes all the extra Higgs bosons massive, reasonably leading to a phenomenologically consistent scalar spectrum as shown below. For partial solutions of the potential minimization and scalar spectrum, let us call for the reader's attention to \cite{pires0,pires}.   

To identify the scalar particles, let us expand the neutral fields (and the conservation of $W$-parity must be maintained as determined above): 
\bea
\rho &=& \left(\begin{array}{c}
\rho^+_1\\
\fr{1}{\sqrt{2}}(v+S_2+i A_2)\\
\rho^+_3\end{array}\right),\hs
\eta =  \left(\begin{array}{c}
\fr{1}{\sqrt{2}}(u+S_1+iA_1)\\
\eta^-_2\\
\fr{1}{\sqrt{2}}(S'_3+iA'_3)\end{array}\right),\hs
 \chi = \left(\begin{array}{c}
\fr{1}{\sqrt{2}}(S'_1+i A'_1)\\
\chi^-_2\\
\fr{1}{\sqrt{2}}(\om + S_3+iA_3)\end{array}\right),\eea where $S'_{1,3}$ and $A'_{1,3}$ are $W$-parity odd while $S_{1,2,3}$ and $A_{1,2,3}$ are even. Two kinds of these particles do not mix. Similarly, for the charged scalars, $\rho_1$ and $\eta_2$ do not mix with $\rho_3$ and $\chi_2$. All these can be seen by the results given below. The potential minimization conditions are obtained by              
 \bea && v \mu^2_1 + v^3 \la_1 + \fr 1 2 v \om^2 \la_4 + \fr 1 2 v u^2 \la_5 +\fr{1}{\sqrt{2}}f u \om=0,\\ 
 && u \mu^2_3 + u^3 \la_3 +\fr 1 2 u v^2 \la_5 +\fr{1}{2} u \om^2 \la_6 + \fr{1}{\sqrt{2}} f v \om=0,\\
 && \om \mu^2_2 +\om^3 \la_2 + \fr 1 2 \om v^2 \la_4 +\fr 1 2 \om u^2 \la_6 +\fr{1}{\sqrt{2}}f u v=0.\eea
 
The pseudoscalars $A_1$, $A_2$ and $A_3$ mix because $f\neq 0$. One combination of these fields is a physical pseudoscalar ($A$) with mass,
 \be A=\fr{u^{-1} A_1+v^{-1}A_2+\om^{-1}A_3}{\sqrt{u^{-2}+v^{-2}+\om^{-2}}},\hs  m^2_A=-\fr{f}{\sqrt{2}}\fr{u^2v^2+u^2\om^2+v^2\om^2}{u v \om}.\ee 
We see that $f<0$ if $u,v,\om>0$. The two other fields are massless, orthogonal to $A$, and identified as the Goldstone bosons of $Z$ and $Z'$ as given by 
 \be G_{Z'} =\fr{-\om^{-1} (u^{-1} A_1+ v^{-1}A_2)+(u^{-2}+v^{-2})A_3}{\sqrt{(u^{-2}+v^{-2}+\om^{-2})(u^{-2}+v^{-2})}},\hs G_Z = \fr{-uA_1+vA_2}{\sqrt{u^{2}+v^{2}}}.\ee  
 The $A$ mass is proportional to $\om$ if $f$ is here (and below) supposed in $\om$ scale ($f\sim -\om$). At the leading order, $\om\gg u,v$, we have $G_{Z'}\simeq A_3$ and $A\simeq (v A_1 + u A_2)/\sqrt{u^2+v^2}$. 
 
 The scalars $S_1$, $S_2$ and $S_3$ mix via the mass Lagrangian:
 \bea \left(S_1\ S_2\ S_3\right)
 \left(\begin{array}{ccc}
 \la_3 u^2-\fr{f v \om}{2\sqrt{2}u} & \fr 1 2 \la_5 u v +\fr{f\om}{2\sqrt{2}}&\fr 1 2 \la_6 u \om + \fr{f v}{2\sqrt{2}}\\
 \fr 1 2 \la_5 u v +\fr{f\om}{2\sqrt{2}} & \la_1 v^2 -\fr{f u \om}{2\sqrt{2}v} & \fr 1 2 \la_4 v \om + \fr{f u }{2\sqrt{2}} \\
 \fr 1 2 \la_6 u \om + \fr{f v}{2\sqrt{2}} &\fr 1 2 \la_4 v \om + \fr{f u }{2\sqrt{2}} & \la_2 \om^2 -\fr{f u v}{2\sqrt{2}\om}
 \end{array}\right)
 \left(\begin{array}{c}
 S_1\\
 S_2\\
 S_3\end{array}\right). \eea This mass matrix always gives a physical light state to be identified as the standard model Higgs boson ($H$). Since $f\sim -\om$, the two other physical states ($H_{1,2}$) are heavy living in $\om$ scale. At the leading order ($-f,\om\gg u,v$), those physical fields with respective masses can be obtained by 
 \bea H_1&\simeq& \fr{-v S_1+uS_2}{\sqrt{u^2+v^2}},\hs m^2_{H_1}\simeq -\fr{f\om}{\sqrt{2}}\left(\fr{u}{v}+\fr{v}{u}\right),\hs H_2\simeq S_3,\hs m^2_{H_2}\simeq 2\la_2\om^2,\\
 H &\simeq& \fr{u S_1+vS_2}{\sqrt{u^2+v^2}},\hs m^2_H\simeq \fr{4\la_2(\la_5u^2v^2+\la_3 u^4+\la_1 v^4)-(\sqrt{2}uv(f/\om)+\la_6 u^2+ \la_4 v^2)^2}{2\la_2(u^2+v^2)}.\eea  

One combination of $S'_1$ and $S'_3$ is a physical field
 $S'=(\om S'_3+uS'_1)/\sqrt{u^2+\om^2}$ with mass $m^2_{S'}=\left(\fr 1 2 \la_9-\fr{f v}{\sqrt{2} u \om}\right)(u^2+\om^2)$. The orthogonal state 
 $G'_S=(-u S'_3+\om S'_1)/\sqrt{u^2+\om^2}$ is a massless Goldstone field. Similarly, one combination of $A'_1$ and $A'_3$ is a physical field
 $A'=(\om A'_3 - u A'_1)/\sqrt{u^2+\om^2}$ with mass $m^2_{A'}=\left(\fr 1 2 \la_9 -\fr{f v}{\sqrt{2}u \om}\right)(u^2+\om^2)$. The orthogonal state $G'_A=(u A'_3 +\om A'_1)/\sqrt{u^2+\om^2}$ is a massless Goldstone field. It is easy to realize that the $G'_S$ and $G'_A$ are Goldstone bosons of $\mathrm{Re}X$ and $\mathrm{Im}X$ gauge fields, respectively. Therefore, their combination can be identified as the Goldstone boson of the $X$ gauge boson: \be G_{X}=\fr{1}{\sqrt{2}}(G'_S+iG'_A)=\fr{\om\chi_1-u\eta^*_3}{\sqrt{u^2+\om^2}}.\ee Simultaneously, we also have a physical neutral complex field as a combination of $S'$ and $A'$ (and obviously orthogonal to $G_X$) with the mass as given: 
 \be H'=\fr{1}{\sqrt{2}}(S'+iA')=\fr{u\chi^*_1+\om\eta_3}{\sqrt{u^2+\om^2}},\hs m^2_{H'}=\left(\fr 1 2 \la_9-\fr{f v}{\sqrt{2} u \om}\right)(u^2+\om^2).\ee The $H'$ is only physical neutral scalar field which is odd under $W$-parity responsible for dark matter as shown below. It is to be noted that the $H'$ mass is always in $\om$ scale. At the leading order ($\om\gg u,v$) we have $H'\simeq \eta_3$ (which is a scalar singlet of the standard model) and $G_X\simeq \chi_1$.    
 
There are two physical charged scalars, one $W$-parity odd ($H_3$) and another even ($H_4$):
 \be H^-_3=\fr{v \chi^-_2+\om\rho^-_3}{\sqrt{v^2+\om^2}},\hs H^-_4=\fr{v \eta^-_2+u\rho^-_1}{\sqrt{u^2+v^2}},\ee with respective masses
 \be m^2_{H_3}=\left(\fr 1 2 \la_7 -\fr{fu}{\sqrt{2}v \om}\right)(v^2+\om^2),\hs m^2_{H_4}=\left(\fr 1 2 \la_8-\fr{f\om}{\sqrt{2}u v}\right)(u^2+v^2).\ee The orthogonal states to these scalars are Goldstone bosons of the $Y$ and $W$ bosons, respectively:
 \be G^-_Y=\fr{\om \chi^-_2-v\rho^-_3}{\sqrt{v^2+\om^2}},\hs G^-_{W}=\fr{u \eta^-_2-v\rho^-_1}{\sqrt{u^2+v^2}}.\ee The masses of $H_3$ and $H_4$ are in $\om$ scale. At the leading order, we have $H_3\simeq \rho_3$ and $G_Y\simeq \chi_2$. 
 
We conclude that only the standard model like Higgs boson $H$ is light in $u,v$ scale. All the other physical scalars such as $A$, $H_{1,2,3,4}$ and $H'$ are heavy in $\om$ scale, while $R$ is in $\La$ scale. The number of the Goldstone bosons match those of the massive gauge bosons (in the case the $U(1)_N$ gauge symmetry is turned on, the extra scalar $\phi$ as required will completely break this charge as well as providing the Goldstone boson $I$ for it). On the other hand, if the $f$ coupling is suppressed as in the literature, the field $A\sim vA_1+u A_2$ becomes a physical massless Goldstone field living in the doublets of the standard model which is very unrealistic. In addition, the identification in \cite{tonasse} in another 3-3-1 version of $\mathrm{Im}\chi^0_3$ (similar to $A_3$ in this model) as a dark matter is incorrect since it is already the Goldstone boson of $Z'$ gauge boson ($G_{Z'}$).        

For the gauge boson sector after integrating out $Z_N$, the masses of the remaining gauge bosons are given as usual:  
\be c^2_W m^2_Z\simeq m^2_W=\fr{g^2}{4}(u^2+v^2),\hs m^2_Y=\fr{g^2}{4}(v^2+\om^2),\hs m^2_X=\fr{g^2}{4}(u^2+\om^2), \ee and $Z'$ obtaining a mass in $\om$ scale as $X$ and $Y$ (to be specified in the next section), where $g$ is $SU(3)_L$ gauge coupling constant. We therefore identify $u^2+v^2=(246\ \mathrm{GeV})^2$. It is noted that $W$ and $Y$ do not mix. Similarly $Z$, $Z'$ and $Z_N$ do not mix with $X$. All these are due to the $W$-parity symmetry that forces the VEVs of $\chi^0_1$ and $\eta^0_3$ vanishing. Moreover, the mass spectrum of neutral gauge boson sector will be changed if $Z_N$ gets a mass in the 3-3-1 scale ($\La\sim \om$). By our convention as given, this should be skipped in the present work.    

For the fermion sector, we first note that the Dirac masses appeared below will be written in the Lagrangian of form $-\bar{f}_L m_f f_R+h.c.$ The masses of exotic quarks are given by \be m_U=-\fr{1}{\sqrt{2}}h^U \om,\hs  [m_{D}]_{\al\beta}=-\fr{1}{\sqrt{2}}h^D_{\al\beta}\om,\ee which are all in $\om$ scale. The ordinary quarks and charged leptons get consistent masses at the tree-level as in the 3-3-1 model with right-handed neutrinos: \be [m_u]_{\al a}=\fr{1}{\sqrt{2}}h^u_{\al a}v,\hs [m_u]_{3a}=-\fr{1}{\sqrt{2}}h^u_{a}u,\ee
for up quarks, \be [m_d]_{\al a}=-\fr{1}{\sqrt{2}}h^d_{\al a}u,\hs [m_d]_{3a}=-\fr{1}{\sqrt{2}}h^d_{a}v,\ee for down quarks, and \be [m_e]_{ab}=-\fr{1}{\sqrt{2}}h^e_{ab} v, \ee for charged leptons. Here, we see that the up quarks do not mix with $U$, and the down quarks also do not mix with $D_\al$ as already mentioned. 

The $\nu_L$ and $\nu_R$ as coupled to $\eta^0_1$ will have Dirac masses:
\be [m^D_\nu]_{ab}=-\fr{1}{\sqrt{2}}h^\nu_{ab} u. \ee However, the right-handed neutrinos $(\nu_R)$ by themselves coupled via $\phi$ will get large Majorana masses (in $\La$ scale) with the form, $-\fr{1}{2} \bar{\nu}^c_R m^M_\nu \nu_R +h.c.,$ where \be [m^M_\nu]_{ab}=-\sqrt{2}h'^\nu_{ab}\La.\ee Consequently, the (observed) active neutrinos ($\sim \nu_L$) get naturally small masses via a type I seesaw mechanism as given by 
\be m^{\mathrm{eff}}_\nu=-m^D_\nu (m^M_\nu)^{-1} (m^D_\nu)^T\sim \fr{(h^\nu)^2}{h'^\nu}\fr{u^2}{\La}. \ee If the $\La$ is proportional to $\om$ acting on TeV scale, the Dirac mass parameters ($m^D_\nu$) should get values in the electron mass range in order for $m^{\mathrm{eff}}_\nu$ in sub eV. In any case, the masses of physical sterile neutrinos $(\sim \nu_R)$ are in $\La$ scale responsible for the $U(1)_N$ breaking.    
         
Unlike the previous model \cite{331r}, the $N_R$ have vanishing masses at the renormalizable level because $\rho$ does not couple to $\psi_L\psi_L$ (in addition, $\chi$ is also not coupled to $\psi_L\nu_R$) due to the $\mathcal{L}$-charge or $U(1)_N$ symmetry. [In the 3-3-1 model with right-handed neutrinos \cite{331r}, the status is not better. Although the $\psi_L\psi_L\rho$ coupling is allowed, the tree-level neutrinos have only three Dirac masses, no Majorana type, in which one mass is zero and two others degenerate that are unrealistic under the data~\cite{pdg}]. Fortunately, the masses of $N_R$ can be generated by the scalar content by itself via an effective operator invariant under the 3-3-1-1 symmetry and $W$-parity:
\be \fr{\la_{ab}}{M}\bar{\psi}^c_{aL}\psi_{bL}(\chi\chi)^*+h.c.\ee There are no other operators of types $\psi\psi\eta\eta$ and $\psi\psi\eta\chi$ as often considered due to the 3-3-1-1 symmetry. Consequently, only the neutral fermions get masses via this kind of interactions:
\be [m_{N_R}]_{ab}=-\la_{ab}\fr{\om^2}{M}.\ee 

The mass scale of $N_R$ is unknown, however it can be taken in TeV order (i.e. $M$ is in or not so high compared to $\om$) due to the following facts: (i) In the 3-3-1 model with neutral fermions, $m_{N_R}$ were proved to be naturally in $\om$ scale (but the $W$-parity should be violated) \cite{dongfla}, (ii) We can introduce a new scalar sextet coupled to $\psi_L\psi_L$ conserving the 3-3-1-1 symmetry and $W$-parity. The 33 component of sextet provides the masses for $N_R$. However, it is also responsible for the 3-3-1 symmetry breaking which should be in the same $\om$ scale. On the other hand, this sextet if included can be also reserved for totally breaking the $N$-charge and obtaining $W$-parity due to the VEV of the 11 component carrying a lepton number of two units \cite{331seesaw,dongfla}. The $\nu_L$ neutrinos also get small masses via a type II seesaw by this case. However, let us neglect the scalar sextet by this work because the scalar singlet $\phi$ as included is just enough for all purposes. Finally, it is noted that due to $W$-parity, $\nu_{L,R}$ do not mix with $N_R$. Also, the mass sources of $N_R$ and $\nu_R$ might come from different kinds of the 3-3-1-1 breaking.  
 
\section{\label{modelrelic} Dark matter abundance and direct detection}

Let us note that all the $W$-particles including the dark matter candidates $X^0$, $N^0_R$ and $H'^0$ are heavy particles with the masses proportional to $\om$. Among the $W$-particles, supposing $X^0$ as a LWP ($m_X<m_{N_R},\ m_{U},\ m_{D_\al},\ m_{H_3},\ m_{H'}$), it will be stabilized responsible for dark matter. Notice that $X$ cannot decay into $Y$ and wise versa. This is the first case discussed below. For the second case, $N_R$ will be assumed as a LWP ($m_{N_R}<m_X,\ m_Y,\ m_{U},\ m_{D_\al}, m_{H_3},\ m_{H'}$) for dark matter. The work in \cite{pires} has presented numerical calculations for relic densities using MicrOMEGAs package. In the following, we will give an analytic evaluation. For the constraints from the direct and indirect detection experiments, let us call for the reader's attention to \cite{pires}. Below, we provide only one of these kinds by analytic calculation so that our conclusions are viable.   

The scalar dark matter candidate $H'$, which behaves as a LWP ($m_{H'}<m_X$, $m_Y$, $m_{U}$, $m_{D_\al}$, $m_{H_3}$, $m_{N_R}$), has been traditionally studied in the 3-3-1 models \cite{longlan}. In the following consideration, this candidate will be neglected. For detailed calculations and experimental constraints, let us recall the reader's attention to Refs. \cite{pires0,pires,scalardm}. However, let us make some remarks on this particle: (i) The previous studies \cite{longlan,pires0,pires} that identify the massive scalar $H'\simeq \eta_3$ as a dark matter are unnatural since the symmetries protecting it from decay are either neglected or included in term of lepton charge, $G$-charge, or even $Z_2$ which must be broken due to the problems as shown above. In this case, it will develop a VEV allowing decay channels into the standard model particles such as $H'\longrightarrow HH$ since this field should be naturally heavy (the finetuning in mass was needed in \cite{longlan} which is very ackword). In our model, by the investigation of $W$-parity, the stability issue of $H'$ has been solved, similar to the standard model extension with a $Z_2$ odd scalar singlet. (ii) $H'\simeq \eta_3$ is a singlet under the standard model symmetry, and it annihilates into the standard model particles via the scalar portal, exotic quarks, and new gauge bosons.            
 
\subsection{Relic density of $X^0$ gauge boson}

\begin{figure}[h]
\begin{center}
\includegraphics{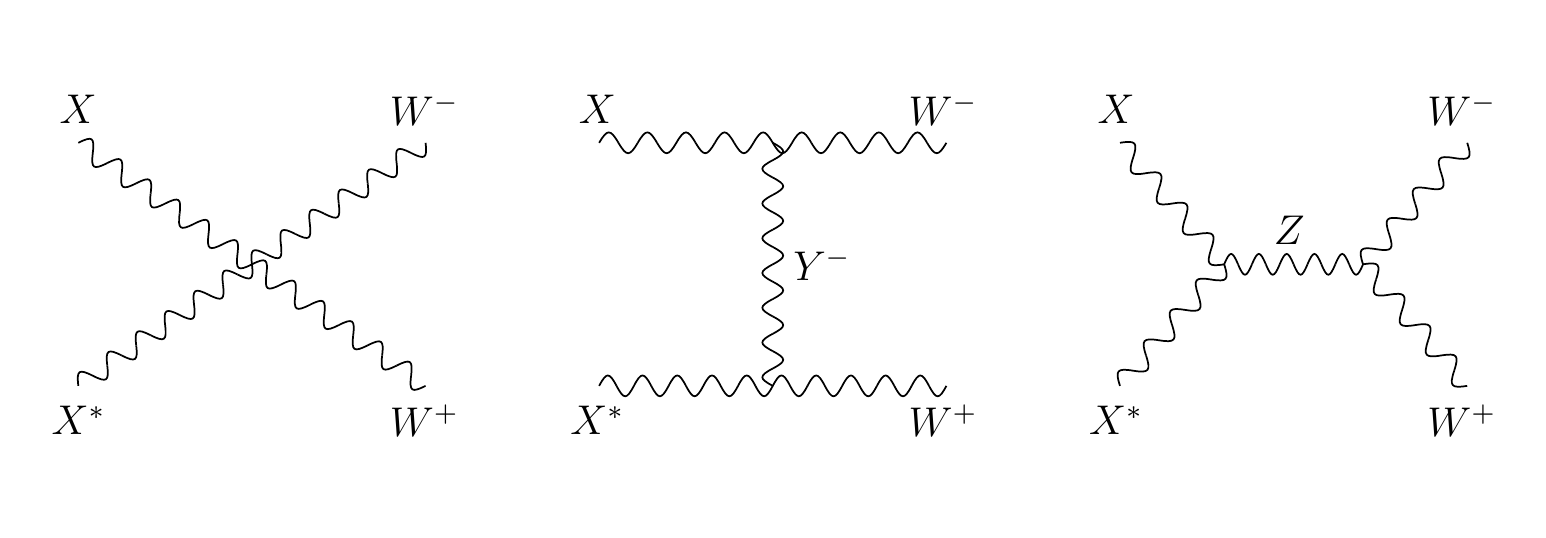}
\caption[]{\label{fig1} Dominant contributions to annihilation of $X^0$ into $W^+W^-$.}
\end{center}
\end{figure}

The annihilation of $X$ into the standard model particles is dominated by the following channels 
\be XX^*\longrightarrow W^+W^-,\ ZZ,\ HH,\ \nu\nu^c,\ l^+l^-,\ qq^c,\ee where $\nu=\nu_e, \nu_\mu, \nu_\tau$, $l=e,\mu,\tau$, and $q=u,d,c,s,t,b$. 
Let us consider the channel $XX^*\rightarrow W^+W^-$ among them. This process is contributed by the diagrams as in Fig. \ref{fig1}.
The Feynman rules can be found in \cite{longsoa}. To the leading order, the thermal average of the cross-section times relative velocity~\cite{reliccal} is given as follows 
\be \langle \sigma v_{\mathrm{rel}}\rangle_{v\rightarrow 0}\simeq \fr{5\al^2 m^2_X}{8s^4_Wm^4_W}. \ee 

Because $m^2_X \gg m^2_W$, this result is too large so that the X can meet the criteria of a dark matter. In fact, the relic density of the candidate \cite{reliccal} is bounded by 
\bea \Omega_X h^2 \simeq \fr{0.1\ \mathrm{pb}}{\langle \sigma_{\mathrm{tot}} v_{\mathrm{rel}}\rangle} <  \fr{0.1\ \mathrm{pb}}{\langle \sigma v_{\mathrm{rel}}\rangle}\simeq 0.0024\times\left(\fr{m_W}{m_X}\right)^2<0.00008,\eea if we take a previous limit on the mass of $X$: $m_X>440$ GeV \cite{toju}. The upper bound of the relic density is too small to compare to the WMAP data $\Omega_{\mathrm{DM}}h^2\simeq 0.11$ \cite{pdg}. The $X$ cannot be a dark matter. This conclusion is coincided with a mere note in \cite{pires}. 

\subsection{Relic density of neutral fermion $N_R$}

\begin{figure}[h]
\begin{center}
\includegraphics{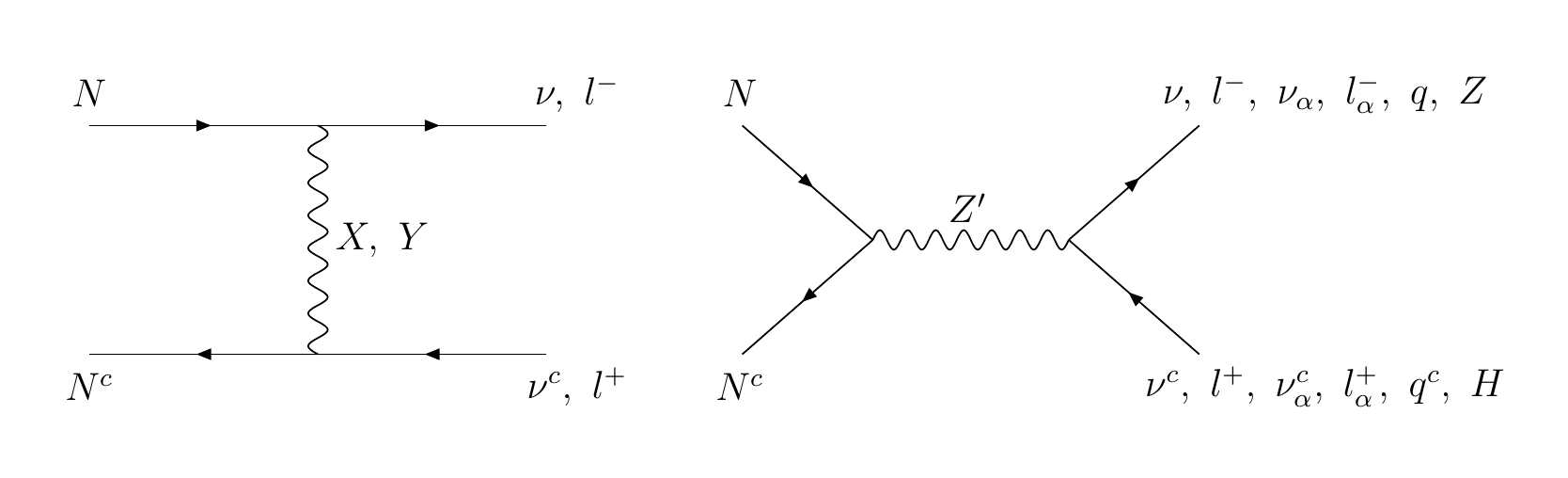}
\caption[]{\label{fig2} Dominant contributions to annihilation of $N_R$.}
\end{center}
\end{figure}

Among the three neutral fermions, $N_{aR}$, assuming $N_R$ is the lightest particle. In addition, supposing $\nu$ and $l$ is the left-handed neutrino and charged lepton that directly couple to $N_R$ via the new gauge bosons $X$ and $Y$, respectively.  There are two other neutrinos and two other charged leptons to be denoted by $\nu_{\al}$ and $l _{\al}$, respectively. The annihilation of $N_R$ into the standard model particles is dominated by the following channels:
\be NN^c\longrightarrow \nu\nu^c,\ l^-l^+,\ \nu_\al \nu^c_\al,\ l^-_\al l^+_\al,\ qq^c,\ ZH,\ee which are given in terms of Feynman diagrams as in Fig. \ref{fig2}.
 Let us remind the reader that there are not channels into $HH$, $W^-W^+$ and $ZZ$ bosons because $\nu_L$ and $N_R$ do not mix due to $W$-parity. 
On the other hand, there may include also some contributions coming from mediated scalars instead of the new gauge bosons, but they are all small and thus neglected.     

The Feynman rules for the above processes can be found in \cite{331r,ninhlong} (see also \cite{ecn331}). An evaluation of the thermal average cross-section times relative velocity is given by 
\bea \langle \sigma v_{\mathrm{rel}}\rangle &\simeq& \fr{g^4m^2_{N_R}}{32\pi}\left(1+\fr{8}{x_F}\right)\left[\fr{1}{m^4_X}+\fr{1}{m^4_Y}-\fr{2c_{2W}}{(3-4s^2_W)m^2_{Z'}} \left(\fr{1}{m^2_X}+\fr{1}{m^2_Y}\right)
\right.\crn &&\left.+\fr{60-104s^2_W+196s^4_W}{3(3-4s^2_W)^2m^4_{Z'}}\right]+\fr{g^4m^2_{N_R}}{32\pi}\sqrt{1-\fr{m^2_t}{m^2_{N_R}}} \left\{\left[1+\fr{1}{x_F}\left(8+\fr{m^2_t}{m^2_{N_R}}\fr{m^2_{N_R}+2m^2_t}{m^2_{N_R}-m^2_t}\right)\right]\right.\crn
&&\times\left. \fr{9-12s^2_W+20s^4_W}{3(3-4s^2_W)^2m^4_{Z'}} +\left[1+\fr{6}{x_F}\left(1+\fr 1 2 \fr{m^2_t}{m^2_{N_R}-m^2_t}\right)\right]\fr{4s^2_W(3-2s^2_W)}{3(3-4s^2_W)^2m^4_{Z'}}\fr{m^2_t}{m^2_{N_R}}\right\}\crn
&&+\fr{g^4m^2_{N_R}}{32\pi}\sqrt{1-\fr{m^2_H+m^2_Z}{2m^2_{N_R}}}\fr{c^2_Wc^2_{2W}}{(3-4s^2_W)^2m^4_{Z'}}\left[1+\fr{2m^2_Z-m^2_H}{2m^2_{N_R}}\right.\crn
&&\left.+\fr{1}{x_F}\left(4+\fr{5m^2_H+11m^2_Z}{2m^2_{N_R}}\right)\right],\eea where we have used the facts that $m_\nu,\ m_l,\ m_q\ (q\neq t)\ll m_t,\ m_Z,\ m_H < m_{N_R} < m_{X},\ m_Y,\ m_{Z'}.$ In addition, the above cross section has been expanded in the non-relativistic limit of $N_R$ as usual up to the squared velocity, in which $\langle v^2\rangle = 6/x_F$ and $x_F\equiv m_{N_R}/T_F\sim 20$ at freeze-out temperature~\cite{reliccal}. 

To have a numerical value for the WMAP data, let us use the condition $\om^2\gg u^2,v^2$ which follows the tree-level relation (the mass of $Z'$ can be found in \cite{dongepjc}), \be m^2_X\simeq m^2_Y\simeq \fr{3-t^2_W}{4} m^2_{Z'}.\ee 
Also let the neutral fermion mass be enough large $m^2_{N_R}\gg m^2_{t,W,H}$ so that the ratios $m^2_{t,H,W}/m^2_{N_R}$ can be neglected and the new physics is safe. We have 
\bea \langle \sigma v _{\mathrm{rel}}\rangle \simeq \fr{\al^2}{(150\ \mathrm{GeV})^2}\fr{(2557.5\ \mathrm{GeV})^2 m^2_{N_R}}{m^4_{Z'}},\eea where we have used $s^2_W=0.231$ and $x_F=20$. Because $\al^2/(150\ \mathrm{GeV})^2\simeq 1 \mathrm{pb}$, the WMAP data on the dark matter relic density for $N_R$ ($\Omega_N h^2\simeq 0.1\mathrm{pb}/\langle \sigma v_{\mathrm{rel}}\rangle\simeq 0.11$) implies 
\be m_{N_R}\simeq \fr{m^2_{Z'}}{2557.5\ \mathrm{GeV}}.\ee Since $m_{N_R}< m_{Z'}$ we derive  $m_{Z'}\leq 2.5\ \mathrm{TeV}$. This upper limit of $Z'$ mass is needed in order to make the dark matter candidate $N_R$ stable. The several lower limits on $Z'$ mass have been given in the literature as some TeV \cite{pdg,masszp}, so let us take the strong one recently studied in the second article of \cite{masszp}, $m_{Z'}\geq 2.2$ TeV. Consequently, the mass of $N_R$ is limited by $m_{N_R}\geq 1.9$ TeV. Summary, the $N_R$ is a dark matter if it has a mass in the range: \be 1.9\ \mathrm{TeV} \leq m_{N_R}\leq 2.5\ \mathrm{TeV}.\ee The mass of $N_R$ is completely fixed via $m_{Z'}$ or the VEV $\om$ as a single-valued function due to the relic density as given which is unlike that in \cite{pires} numerically calculated with the MicrOMEGAs package. Our result above is in agreement with the large range among others as dedicated in \cite{pires}.  

\subsection{Direct detection of dark matter $N_R$} 

The direct detection experiments measure the recoil energy deposited by the scattering of dark matter with the nuclei in a large detector. At the fundamental level, the scattering is due to the interactions of dark matter with quarks as confined in the nucleons. In this model, the leading contribution to the $N_R$-quark scattering amplitude comes from the t-chanel exchange of $Z'$ boson (there may be another contribution of $Z$ boson, however it is very suppressed due to the contrained small mixing of $Z-Z'$). Therefore, the effective Lagrangian is given by 
\be \mathcal{L}^{\mathrm{eff}}_{N_R-\mathrm{quark}}=\bar{N}_R\ga^\mu N_R [\bar{q}\ga_\mu (\al_q P_L+\beta_q P_R)q],\ee where the relevant couplings are evaluated by \cite{331r,ninhlong,ecn331} 
\be \al_{u,d,c,s}=-\fr{g^2}{6m^2_{Z'}},\hs \beta_{u,c}=\fr{2g^2s^2_W}{3(4c^2_W-1)m^2_{Z'}},\hs \beta_{d,s}=-\fr{g^2s^2_W}{3(4c^2_W-1)m^2_{Z'}}.\ee
In the non-relativistic limit, there are only two operators in the effective Lagrangian surviving (the other operators vanish) as given by \cite{bbpsdd}
\be \mathcal{L}^{\mathrm{eff}}_{N_R-\mathrm{quark}}=\la_{q,o} \bar{N}\ga^\mu N \bar{q}\ga_\mu q  + \la_{q,e} \bar{N}\ga^\mu \ga_5 N \bar{q}\ga_\mu\ga_5 q,\ee where $\la_{q,o}\equiv (\beta_q+\al_q)/4$ is of the odd operator while $\la_{q,e}\equiv (\beta_q-\al_q)/4$ for the even operator.     

The $N_R$-nucleon amplitudes can be directly converted from the amplitudes above via the nucleon form factors as obtained by \cite{bbpsdd}
\be \mathcal{L}^{\mathrm{eff}}_{N_R-{\mathrm{nucleon}}}=\la_{\psi,o}\bar{N}\ga^\mu N \bar{\psi}\ga_\mu \psi+\la_{\psi,e}\bar{N}\ga^\mu\ga_5 N \bar{\psi}\ga_\mu\ga_5 \psi, \ee
where $\psi$ is nucleon, $\psi\equiv (p,n)$, and $\la_{\psi,e}=\sum_{q=u,d,s}\Delta^\psi_q \la_{q,e}$ with the $\Delta^\psi_q$ values as provided in \cite{bbpsdd}, while $\la_{\psi,o}$ is given by \be \la_{\psi,o}=\sum_{q=u,d}f^\psi_{Vq} \la_{q,o},\hs f^p_{Vu}=2,\hs f^p_{Vd}=1,\hs f^n_{Vu}=1,\hs f^n_{Vd}=2.\ee 
The $\la_{\psi,o}$ and $\la_{\psi,e}$ are spin-independent (SI) and spin-dependent (SD) interactions, respectively. 

For the large neuclei, the $N_R$-nucleus scattering cross-section is strongly enhanced due to the SI interaction, while there is no strong enhancement from the SD interaction \cite{bbpsdd}. Therefore, the dominant contribution to the cross-section comes from the SI interaction as given by
\be \sigma^{\mathrm{SI}}_{N_R-\mathrm{nucleus}}=\fr{4\mu^2_{A}}{\pi}(\la_p Z + \la_n (A-Z))^2, \ee where $Z$ is the nucleus charge, $A$ the total number of nucleons, and \be \mu_{A}=\fr{m_{N_R}m_A}{m_{N_R}+m_A},\hs \la_p = \fr{\la_{p,o}}{2}=\fr{3(2s^2_W-1)g^2}{16(3-4s^2_W)m^2_{Z'}},\hs \la_n=\fr{\la_{n,o}}{2}=-\fr{g^2}{16m^2_{Z'}}.\ee  The experimental output for the $N_R$-nucleon cross-section is the above result per a nucleon, 
\be \sigma^{\mathrm{SI}}_{N_R-\mathrm{nucleon}}=\fr{4\mu^2_{\mathrm{nucleon}}}{\pi}\left(\la_p\fr{Z}{A}+\la_n\fr{A-Z}{A}\right)^2,\hs \mu_{\mathrm{nucleon}}=\fr{m_{N_R}m_{p,n}}{m_{N_R}+m_{p,n}}\simeq m_{\mathrm{nucleon}}.\ee             

The strongest limit on the $N_R$-nucleon cross-section presently comes from the XENON100 experiment. Taking the Xe nucleon with $Z=54,\ A=131$, and $m_{\mathrm{nucleon}}\simeq 1$ GeV, $g^2=4\pi\al/s^2_W$ with $\al=1/128$ and $s^2_W=0.231$, we have
\be \sigma^{\mathrm{SI}}_{N_R-\mathrm{nucleon}}\simeq 2.9\times 10^{-43} \left(\fr{1\ \mathrm{TeV}}{m_{Z'}}\right)^4\ \mathrm{cm}^2.\ee With the $Z'$ mass limit as given above, $m_{Z'}\geq 2.2$ TeV, it leads to \be \sigma^{\mathrm{SI}}_{N_R-\mathrm{nucleon}}\leq 1.2\times 10^{-44}\ \mathrm{cm}^2.\ee This limit is in good agreement to the constraint from the XENON100 experiment \cite{xenon100} since our dark matter is heavy with the mass in TeV range.

\section{\label{con}Conclusion and outlook}

We have given a detailed analysis of lepton and baryon numbers in the 3-3-1 model with neutral fermions. If they are (residual) symmetries of the theory, which are strictly respected by the gauge interactions, minimal Yukawa Lagrangian and scalar potential, they behave as local symmetries. We have also given a classification of the wrong-lepton particles that have anomalous lepton (even baryon) numbers, whereas the ordinary particles including the standard model ones do not have this property. Moreover, all the unwanted interactions, which lead to the tree-level flavor changing neutral currents, inconsistent neutrino masses, instability of the lightest wrong-lepton particle, and further the tree-level CPT violation, are naturally suppressed due to one of those symmetries. It is generally applied also for the 3-3-1 model with right-handed neutrinos and the minimal 3-3-1 model if these theories conserve the lepton and baryon number symmetries.  

The lepton ($\mathcal{L}$) and baryon ($\mathcal{B}$) numbers can be as several appearances of and unified in a single charge ($N=\mathcal{B-L}$) of natural gauge symmetry $SU(3)_C\otimes SU(3)_L\otimes U(1)_X\otimes U(1)_N$ independent of all the anomalies such as leptonic and baryonic, recognizing $B-L=-(2/\sqrt{3})T_8+N$ as a charge of $SU(3)_L\otimes U(1)_N$ (see also \cite{sendixit} for another 3-3-1-1 symmetry derived from grand unifications). All the unwanted interactions is not invariant under this gauge symmetry which is prevented, while the minimal Lagrangian of the 3-3-1 theory respects it. A direct consequence of the 3-3-1-1 model is that it  contains by itself a residual discrete symmetry after breaking, $W$-parity: $P_W=(-1)^{-2\sqrt{3}T_8+3N+2s}$, as a nature symmetry of the wrong-lepton particles. The unwanted vacuums, which also lead to the problems as mentioned above, are naturally discarded due to this parity. The lightest wrong-lepton particle is truly stabilized due to the 3-3-1-1 gauge symmetry with unbroken $W$-parity responsible for the dark matter. With the aid of $W$-parity, it is completely understood why the wrong-lepton particles are only coupled in pairs in the minimal Lagrangian with the standard vacuum structure, due to the specific 3-3-1 gauge symmetry (which has been explicitly shown in the text too). This $W$-parity has a natural origin of the 3-3-1-1 gauge symmetry by itself which is discriminated from those in the MSSM and others with gauged $B-L$ \cite{lrm,so10}.

We have also provided a general analysis of the scalar sector, identifying physical scalars and Goldstone bosons. The trilinear coupling of scalars should present in order to make all extra Higgs bosons massive, keeping the model obviously consistent with the low energy theory. The pseudoscalar part of $\chi_3$ should be the Goldstone boson of $Z'$ gauge boson which is unlike the conclusion in \cite{tonassescalar}. Therefore, it should be not a dark matter \cite{tonasse}. Finally, this sector will be charged if the $U(1)_N$ gauge symmetry is turned on.              

We have explicitly shown that the non-Hermitian neutral gauge boson ($X$) cannot be a dark matter. However, the neutral fermion ($N_R$) can contribute the dark matter if its mass is given in the range $1.9\ \mathrm{TeV}\leq m_{N_R}\leq 2.5\ \mathrm{TeV}$, provided that the mass of $Z'$ gauge boson satisfying $2.2\ \mathrm{TeV}\leq m_{Z'}\leq 2.5\ \mathrm{TeV}$. In these calculations, we have neglected the contributions of the new gauge boson $Z_N$ (it should be assumed to be so massive or weakly interacting). If its mass and coupling are comparable to those of $X,\ Y,\ Z'$, our results may be changed. Also, phenomenologies in the 3-3-1-1 model such as the baryon number asymmetry, neutrino masses, and new physics associated with the $Z_N$ gauge boson will be very interesting. All these and the above one are devoted to further studies to be published elsewhere \cite{dongbl331}.                        

The $W$-parity transforms trivially in the 3-3-1 model with right-handed neutrinos, i.e. all the particles in the model is even under this parity. While the model might right predict potential dark matter candidates, let us ask what is the mechanism other than that useless $W$-parity for stabilizing the dark matter. Back to the past \cite{pires0}, the first notes were assigned for the conservation of lepton number and $Z_2$. However, this $Z_2$ is really broken by the VEVs of scalars while the lepton number should be also violated by five dimensional effective interactions or broken by the vacuum responsible for the neutrino masses. On the other hand, the lepton number respected as a symmetry of the theory is also broken as a gauge symmetry and anomalous. In \cite{pires}, the lepton content was changed and the $U(1)_G$ included perhaps to avoid some of those problems. However, this $U(1)_G$ takes the same status as the lepton number that acts as a gauge symmetry and also broken. Fortunately, by the new lepton content this charge is anomaly free like $N=\mathcal{B-L}$. So, we propose one solution to the stability of the dark matter in their model is by imposing $G$-parity ($P_G=(-1)^G$) which is odd for the $G$-particles and even for ordinary particles. In this case, the gauge symmetry should be $SU(3)_C\otimes SU(3)_L\otimes U(1)_X\otimes U(1)_{\mathcal{G}}$ where $G=\fr{2}{\sqrt{3}}T_8+\mathcal{G}$. Now let us turn to the 3-3-1 model with right-handed neutrinos. The $U(1)_G$ is also useless as the lepton number or $W$-parity since it also yields anomalies. To go over all these difficulties, in the following, we suppose a new mechanism based on the idea of a potential ``inert'' scalar triplet naturally realized by a $Z_2$ symmetry with the base of the economical 3-3-1 model \cite{ecn331}. The $W$-parity is explicitly violated in this model, and the lepton number is no longer to be regarded as a gauge symmetry since it is only an approximate symmetry, explicitly violated by the Yukawa interactions.   

We know that the 3-3-1 model with right-handed neutrinos can work with three scalar triplets $(\rho,\ \eta,\ \chi)$ in which two of them $(\eta,\ \chi)$ have the same gauge symmetry quantum numbers \cite{331r}. If we exclude one of these two triplets (assumed $\eta$) it results a new, consistent, predictive model, named the economical 3-3-1 model, working with only the two scalar triplets $(\rho,\ \chi)$ as recently investigated in a series of articles \cite{ecn331,dhhl,dls1,dls2,dongaxion} (note that in those works $\rho$ called as $\phi$ instead). Alternative to that proposal, we can retain $\eta$, but introducing a $Z_2$ symmetry so that the $\eta$ is odd, while the $\chi,\ \rho$ and all other fields are even. The resulting model will be very rich in phenomenology other than the economical 3-3-1 model due to the contribution of $\eta$. In fact, the vacuum can also be conserved by the $Z_2$ as a partial solution of the potential minimization, $\langle \eta \rangle =0$, $\langle \rho \rangle = \fr{1}{\sqrt{2}}(0,v,0)$, and $\langle \chi \rangle =\fr{1}{\sqrt{2}}(u,0,\om)$. We have thus a new economical 3-3-1 model with an inert scalar triplet $\eta$ that is odd under the exact and unbroken $Z_2$ symmetry. The scalar triplets $\chi$ and $\rho$ can break the gauge symmetry and generating the masses for the particles in a correct way like the economical 3-3-1 model \cite{ecn331}. The inert scalar triplet $\eta$ can provide some dark matter candidates, however they may belong to a scalar singlet or a scalar doublet under the standard model symmetry. In the sense the model proposed is quite similar to the two Higgs doublet model in which one doublet is inert, well-known as the inert doublet model \cite{mait}. However, this theory provides only a doublet dark matter. The proposal is to be published elsewhere \cite{dong331it}.

\section*{Acknowledgments}
This research is funded by Vietnam National Foundation for
Science and Technology Development (NAFOSTED) under grant number 103.03-2011.35. PVD would like to thank Dr. Do Thi Huong at Institute of Physics, Vietnam Academy of Science and Technology for useful discussions. 

\appendix

\section{\label{appendixa}Checking the $U(1)_N$ anomalies}

The nontrivial anomalies as associated with $U(1)_N$ that are potentially troublesome can be listed as follows: $[SU(3)_C]^2U(1)_N$, $[SU(3)_L]^2U(1)_N$, $[U(1)_X]^2U(1)_N$, $U(1)_X[U(1)_N]^2$, $[U(1)_N]^3$, and $[\mathrm{Gravity}]^2U(1)_N$. The other anomalies associated with the usual 3-3-1 symmetry obviously vanish \cite{anoma} which will not be considered in this appendix.

With the fermion content and $N$-charges of the 3-3-1-1 model as given in Table \ref{bangn}, the mentioned anomalies can be calculated. The first one is proportional to \bea [SU(3)_C]^2U(1)_N&\sim& \sum_{\mathrm{all\ quarks}} (N_{q_L}-N_{q_R})\crn &=&3N_{Q_3}+2\times 3 N_{Q_\al}-3N_{u_a}-3N_{d_a}-N_U-2N_{D_\al}\crn
&=&3(2/3)+6(0)-3(1/3)-3(1/3)-(4/3)-2(-2/3)=0,\eea which vanishes. The second anomaly also vanishes, 
\bea [SU(3)_L]^2U(1)_N&\sim& \sum_{\mathrm{all\ (anti)triplets}} N_{F_L}\crn
&=&3N_{\psi_a}+3N_{Q_3}+2\times 3 N_{Q_\al}\crn
&=&3(-2/3)+3(2/3)+6(0)=0.\eea Here the number of fundamental colors (the 3s in the second and last terms) must be taken into account. In the following the appearance of color numbers should be understood. Notice also that the relation $\mathrm{Tr}[(-T^*_i)(-T^*_j)N]=\mathrm{Tr}[T_iT_jN]$ has been used. 

The third anomaly is given by 
\bea [U(1)_X]^2U(1)_N&=&\sum_{\mathrm{all\ fermions}}(X^2_{f_L}N_{f_L}-X^2_{f_R}N_{f_R})\crn
&=&3\times 3 X^2_{\psi_a}N_{\psi_a}+3\times 3 X^2_{Q_3} N_{Q_3}+2\times 3\times 3 X^2_{Q_{\al}}N_{Q_{\al}}\crn
&&-3\times 3 X^2_{u_a}N_{u_a}-3\times 3 X^2_{d_a}N_{d_a}-3X^2_U N_U-2\times 3 X^2_{D_\al}N_{D_\al}\crn
&&-3X^2_{e_a} N_{e_a}-3X^2_{\nu_a}N_{\nu_a}\crn
&=&3\times 3 (-1/3)^2(-2/3)+3\times 3 (1/3)^2(2/3)+2\times 3\times 3 (0)^2(0)\crn
&&-3\times3(2/3)^2(1/3)-3\times 3(-1/3)^2(1/3)-3(2/3)^2(4/3)\crn
&&-2\times3(-1/3)^2(-2/3)-3(-1)^2(-1)-3(0)^2(-1)^2=0. \eea The fourth anomaly is similarly calculated,           
\bea U(1)_X[U(1)_N]^2&=&\sum_{\mathrm{all\ fermions}}(X_{f_L}N^2_{f_L}-X_{f_R}N^2_{f_R})\crn
&=&3\times 3 X_{\psi_a}N^2_{\psi_a}+3\times 3 X_{Q_3} N^2_{Q_3}+2\times 3\times 3 X_{Q_{\al}}N^2_{Q_{\al}}\crn
&&-3\times 3 X_{u_a}N^2_{u_a}-3\times 3 X_{d_a}N^2_{d_a}-3X_U N^2_U-2\times 3 X_{D_\al}N^2_{D_\al}\crn
&&-3X_{e_a} N^2_{e_a}-3X_{\nu_a}N^2_{\nu_a}\crn
&=&3\times 3 (-1/3)(-2/3)^2+3\times 3 (1/3)(2/3)^2+2\times 3\times 3 (0)(0)^2\crn
&&-3\times3(2/3)(1/3)^2-3\times 3(-1/3)(1/3)^2-3(2/3)(4/3)^2\crn
&&-2\times3(-1/3)(-2/3)^2-3(-1)(-1)^2-3(0)(-1)^2=0. \eea 

The $U(1)_N$ self-anomaly is 
\bea [U(1)_N]^3&=&\sum_{\mathrm{all\ fermions}}(N^3_{f_L}-N^3_{f_R})\crn
&=&3\times 3 N^3_{\psi_a}+3\times 3 N^3_{Q_3}+2\times 3\times 3 N^3_{Q_\al}\crn
&&-3\times 3 N^3_{u_a}-3\times 3 N^3_{d_a}-3N^3_U-2\times 3 N^3_{D_\al}-3N^3_{e_a}-3N^3_{\nu_a}\crn
&=&3\times 3 (-2/3)^3+3\times 3 (2/3)^3+2\times 3\times 3 (0)^3\crn
&&-3\times3(1/3)^3-3\times 3(1/3)^3-3(4/3)^3-2\times3(-2/3)^3\crn
&&-3(-1)^3-3(-1)^3=0.\eea  The last anomaly is given by
\bea [\mathrm{Gravity}]^2U(1)_N&\sim&\sum_{\mathrm{all\ fermions}}(N_{f_L}-N_{f_R})\crn
&=&3\times 3 N_{\psi_a}+3\times 3 N_{Q_3}+2\times 3 \times 3 N_{Q_\al}\crn
&&-3\times 3 N_{u_a}-3\times 3 N_{d_a}-3N_U-2\times 3 N_{D_\al}-3N_{e_a}-3N_{\nu_a}\crn
&=&3\times 3 (-2/3)+3\times 3 (2/3)+2\times 3\times 3 (0)\crn
&&-3\times 3 (1/3)-3\times 3 (1/3) -3(4/3)-2\times 3 (-2/3)\crn
&&-3(-1)-3(-1)=0.\eea These anomalies are only canceled when the right-handed neutrinos are included, in similarity to the standard model extensions with gauged $B-L$. Indeed, since $B-L=-(2/\sqrt{3})T_8+N$ and the $T_8$ obviously independent of anomalies, the cancellation of $N$ anomalies is equivalent to that of $B-L$. It is noted that the 3-3-1 model with right-handed neutrinos is always free from the $U(1)_N$ anomalies due to its fermion content by itself, while the minimal 3-3-1 model like our case is not. Also, if $U(1)_N$ is imposed in the model of \cite{pires}, it is also not free from the gravitational anomaly.        

\section{\label{appendixb}Derivation of $W$-parity}
The $SU(3)_L\otimes U(1)_N$ symmetry is broken down to $U(1)_{B-L}$ by the VEVs of $\eta$, $\rho$ and $\chi$ because the charge $B-L=-(2/\sqrt{3})T_8+N$ anihinates these vacuums: \be (B-L)\langle \eta\rangle = 0,\hs (B-L)\langle \rho \rangle =0,\hs (B-L)\langle \chi\rangle =0.\ee This is the first stage of symmetry breaking. In the second stage the $B-L$ will be broken. And, this is achieved by the VEV of $\phi$ since \be 
(B-L)\langle \phi\rangle \neq 0.\ee It is to be noted that the $\phi$ VEV also breaks $U(1)_N$ by the first stage. Therefore, the $\phi$ vacuum breaks the $N$ charge totally.  

Now, let us find an unbroken residual symmetry as a discrete subgroup of $U(1)_{B-L}$ [exactly of $SU(3)_L\otimes U(1)_N$]. It must satisfy the following condition:
\be e^{i\al (B-L)}\langle \phi\rangle = \langle \phi \rangle,\ee where $\al$ is a parameter of the $U(1)_{B-L}$ Lie group. Because $B(\phi)=0$, $L(\phi)=-2$, and $\langle \phi\rangle = (1/\sqrt{2})\La \neq 0$, we have \be e^{i2\al}=1=e^{i2k\pi}\Longleftrightarrow \al = k\pi,\hs k=0,\pm1,\pm2,\cdots \ee The subgroup that is conserved by the $\phi$ vacuum is obtained by the elements
\be e^{i\al(B-L)}=e^{i k \pi (B-L)}=(-1)^{k(B-L)}=\{1,(-1)^{3(B-L)}\}, \ee which is exactly a $Z_2$ symmetry. When included the spin symmetry $(-1)^{2s}$, we have \be P=(-1)^{3(B-L)+2s}\ee as an exact, unbroken parity symmetry responsible for $W$-particles in the 3-3-1-1 model since
\bea P|\mathrm{wrong\ lepton\ particle}>&=&-|\mathrm{wrong\ lepton\ particle}>,\crn 
P|\mathrm{ordinary\ or\ bilepton\ particle}>&=&+|\mathrm{ordinary\ or\ bilepton\ particle}>.\eea It is noted that although $B-L$ is generally broken by the vacuum, it leaves a residual $Z_2$ symmetry invariant as realized by the scalar singlet $\phi$. Let us remark that if one includes a scalar sextet as mentioned in the text it also yields the $W$-parity as expected.

Finally, let us remind the reader that all the fields which develop VEVs as given are even under $W$-parity: $\phi^0,\ \eta^0_1,\ \rho^0_2,\ \chi^0_3 \longrightarrow \phi^0,\ \eta^0_1,\ \rho^0_2,\ \chi^0_3$, respectively. However, the $W$-fields $\eta^0_3$ and $\chi^0_1$ are odd, $\eta^0_3\longrightarrow - \eta^0_3$ and $\chi^0_1\longrightarrow -\chi^0_1$, which follow vanishing VEVs. The stability of LWP is a consequence of $W$-parity conservation.

In the minimal 3-3-1 model and the 3-3-1 model with right-handed neutrinos (even the model of \cite{pires}), all the new particles are either ordinary or bilepton. Therefore, even $W$-parity is derived in those models, there is no any particle which is odd under the parity.


\begin{thebibliography}{99}

\bibitem{pdg} J. Beringer {\it et al.} (Particle Data Group), Phys. Rev. D {\bf 86}, 010001 (2012).

\bibitem{seesaw1}
P. Minkowski, Phys. lett. B {\bf 67}, 421 (1977); M. Gell-Mann, P.
Ramond and R. Slansky, {\it Complex spinors and unified theories},
in {\it Supergravity}, edited by P. van Nieuwenhuizen and D. Z.
Freedman (North Holland, Amsterdam, 1979), p. 315; T. Yanagida, in
{\it Proceedings of the Workshop on the Unified Theory and the
Baryon Number in the Universe}, edited by O. Sawada and A.
Sugamoto (KEK, Tsukuba, Japan, 1979), p. 95; S. L. Glashow, {\it
The future of elementary particle physics}, in {\it Proceedings of
the 1979 Carg{\`e}se Summer Institute on Quarks and Leptons},
edited by M. L{\'e}vy et al. (Plenum Press, New York, 1980), pp.
687-713; R. N. Mohapatra and G. Senjanovi{\'c}, Phys. Rev. Lett.
{\bf 44}, 912 (1980).

\bibitem{lrm} J. C. Pati and A. Salam, Phys. Rev. D {\bf 10}, 275
(1974); R. N. Mohapatra and J. C. Pati, Phys. Rev. D {\bf 11},
566, 2558 (1975); G. Senjanovi{\'c} and R. N. Mohapatra, Phys.
Rev. D {\bf 12}, 1502 (1975).

\bibitem{so10} H. Georgi, in {\it Particles and Fields},
edited by C. E. Carlson (A.I.P., New York, 1975); H. Fritzsch and
P. Minkowski, Ann. Phys. {\bf 93}, 193 (1975).

\bibitem{leptog} M. Fukugita and T. Yanagida, Phys. Lett. B {\bf 174},
45 (1986).

\bibitem{331r} M. Singer, J. W. F. Valle and J. Schechter, Phys.
Rev. D {\bf 22}, 738 (1980); J. C. Montero, F. Pisano and V.
Pleitez, Phys. Rev. D {\bf 47}, 2918 (1993); R. Foot, H. N. Long
and Tuan A. Tran, Phys. Rev. D {\bf 50}, 34(R) (1994); H. N. Long,
Phys. Rev. D {\bf 53}, 437 (1996); {\bf 54}, 4691 (1996).

\bibitem{dongfla} P. V. Dong, L. T. Hue, H. N. Long
and D. V. Soa, Phys. Rev. D {\bf 81}, 053004 (2010); P. V. Dong, H. N. Long, D. V. Soa, and V. V. Vien, Eur.
Phys. J. C \textbf{71}, 1544 (2011); P. V. Dong, H. N. Long, C. H. Nam, and V. V. Vien, Phys. Rev. D {\bf 85}, 053001 (2012).

\bibitem{331m} F. Pisano and V. Pleitez, Phys. Rev.  D {\bf 46}, 410 (1992);
P. H. Frampton, Phys. Rev. Lett. {\bf 69}, 2889 (1992); R. Foot,
O. F. Hernandez, F. Pisano and V. Pleitez, Phys. Rev. D {\bf 47},
4158 (1993).

\bibitem{anoma} See P. H. Frampton in \cite{331m}.

\bibitem{longvan} H. N. Long and V. T. Van, J. Phys. G {\bf 25}, 2319 (1999).

\bibitem{ecq} F. Pisano, Mod. Phys. Lett A {\bf 11}, 2639 (1996);
A. Doff and F. Pisano, Mod. Phys. Lett. A {\bf 14},
1133 (1999); C. A. de S. Pires and O. P. Ravinez, Phys. Rev. D
{\bf 58}, 035008 (1998); C. A. de S. Pires, Phys. Rev. D {\bf 60},
075013 (1999); P. V. Dong and H. N. Long, Int. J. Mod. Phys. A
{\bf 21}, 6677 (2006).

\bibitem{pires0} C. A. de S. Pires and P. S. Rodrigues da Silva, JCAP {\bf 0712}, 012 (2007).

\bibitem{pires} J. K. Mizukoshi, C. A. de S. Pires, F. S. Queiroz, and P. S. Rodrigues da Silva,  Phys. Rev. D {\bf 83}, 065024 (2011); J. D. Ruiz-Alvarez, C. A. de S. Pires, F. S. Queiroz, D. Restrepo, and P. S. Rodrigues da Silva, Phys. Rev. D {\bf 86}, 075011 (2012).

\bibitem{tonasse} D. Fregolente and M. D. Tonasse, Phys. Lett. B {\bf 555}, 7 (2003).
 
\bibitem{longlan} H. N. Long and N. Q. Lan, Europhys. Lett. {\bf 64}, 571 (2003); S. Filippi, W. A. Ponce, and L. A. Sanches, Europhys. Lett. {\bf 73},
142 (2006). 

\bibitem{lepto331} M. B. Tully and G. C. Joshi, Phys. Rev. D {\bf 64}, 011301 (2001); D. Chang and H. N. Long, Phys. Rev. D {\bf 73}, 053006 (2006).

\bibitem{ma} E. Ma, Phys. Rev. Lett. {\bf 86}, 2502 (2001).

\bibitem{331seesaw} P. V. Dong and H. N. Long, Phys. Rev. D {\bf 77}, 057302 (2008) [arXiv:0801.4196v1 [hep-ph]]. 

\bibitem{ma1} E. Ma and G. Rajasekaran, Phys. Rev. D {\bf 64}, 113012 (2001).

\bibitem{dongbl331} P. V. Dong, D. T. Huong, and N. T. Thuy, ``The 3-3-1-1 model of electroweak and $B-L$ interactions'', in preparation. 

\bibitem{longsoa} H. N. Long and D. V. Soa, Nucl. Phys. B {\bf 601}, 361 (2001); D. T. Binh, D. T. Huong, T. T. Huong, H. N. Long, and D. V. Soa, J. Phys. G {\bf 29}, 1213 (2003).

\bibitem{huyendl} P. V. Dong, V. T. N. Huyen, H. N. Long, and H. V. Thuy, Advances in High Energy Physics {\bf 2012}, 715038 (2012).

\bibitem{longsca} H. N. Long, Mod. Phys. Lett. A {\bf 13}, 1865 (1998).

\bibitem{scalardm} For general discussions, see also: C. Boehm and P. Fayet, Nucl. Phys. B {\bf 683}, 219 (2004). 
 
\bibitem{reliccal} G. Bertone, D. Hooper, and J. Silk, Phys. Rep. {\bf 405}, 279 (2005); G. Jungman, M. Kamionkowski, and K. Griest, Phys. Rep. {\bf 267}, 195 (1996).

\bibitem{toju} M. B. Tully and G. C. Joshi, Phys. Lett. B {\bf 466}, 333 (1999); Int. J. Mod. Phys. A {\bf 13}, 5593 (1998).

\bibitem{ninhlong} H. N. Long and L. D. Ninh, Phys. Rev. D {\bf 72}, 075004 (2005).

\bibitem{ecn331} W. A. Ponce, Y. Giraldo
and L. A. Sanchez, Phys. Rev. D {\bf 67}, 075001 (2003); P. V.
Dong, H. N. Long, D. T. Nhung and D. V. Soa, Phys. Rev. D {\bf
73}, 035004 (2006); P. V. Dong and H. N. Long, Adv. High Energy
Phys. {\bf 2008}, 739492 (2008).

\bibitem{dongepjc}  P. V. Dong and H. N. Long, Eur. Phys. J. C {\bf 42}, 325 (2005).   

\bibitem{masszp} See, for examples, D. A. Gutierrez, W. A. Ponce, and L. A. Sanchez,  Eur. Phys. J. C {\bf 46}, 497 (2006); Y. A. Coutinho, V. S. Guimaraes, and A. A. Nepomuceno,  arXiv:1304.7907 [hep-ph].

\bibitem{bbpsdd} G. Belanger, F. Boudjema, A. Pukhov, and A. Semenov, Comput. Phys. Commun. {\bf 180}, 747 (2009) [arXiv:0803.2360 [hep-ph]]. 

\bibitem{xenon100} E. Aprile {\it et al.} (XENON100 Collaboration), Phys. Rev. Lett. {\bf 109}, 181301 (2012).

\bibitem{sendixit} S. Sen and A. Dixit, Phys. Rev. D {\bf 71}, 035009 (2005); arXiv:hep-ph/0609277. 

\bibitem{tonassescalar} M. D. Tonasse, Phys. Lett. B {\bf 381}, 191 (1996). 

\bibitem{dhhl} P. V. Dong, Tr. T. Huong, D. T. Huong, and H. N. Long,
Phys. Rev. D {\bf 74}, 053003 (2006).

\bibitem{dls1} P. V. Dong, H. N. Long, and D. V. Soa, Phys. Rev. D {\bf 73}, 075005 (2006).

\bibitem{dls2} P. V. Dong, H. N. Long, and D. V. Soa, Phys. Rev. D {\bf 75}, 073006 (2007).

\bibitem{dongaxion} P. V. Dong, H. T. Hung, and H. N. Long, Phys. Rev. D {\bf 86}, 033002 (2012).

\bibitem{mait} N. G. Deshpande and E. Ma, Phys. Rev. D {\bf 18}, 2574 (1978). 

\bibitem{dong331it} P. V. Dong, ``The 3-3-1 model with inert scalar triplet'', in preparation. 

\end{thebibliography}
\end{document}